%% file: A_Functional_Complexity_Framework_for_the_Analysis_of_Telecommunication_Networks.tex
\documentclass{jssc}

\usepackage[T1]{fontenc}% optional T1 font encoding
\usepackage{tikz}
\usepackage{float}
\usepackage{graphicx}
\usepackage{color,soul}
\usepackage[nolist,nohyperlinks]{acronym}
\usepackage{xcolor}
\colorlet{activecolor}{black}
\colorlet{lastUpdate}{black}
\usepackage{caption}

\def\textsubscript#1%
{$_{\text{#1}}$}

\def\cdd{\mbox{\boldmath$\cdot$}~}
\def\dfrac{\displaystyle\frac}
\input{amssym.def}
\include{graphix}

\input jsscN.tex
\begin{document}

%*************************************************************************************************************
% \biaoti{THE CAPITALIZED TITLE OF YOUR ARTICLE$^*$}{The list of authors' names with the LAST NAME capitalized
% and the authors' names should be separated by "\cdd"}{the first author's name \\ the first author's affiliation
% and Email address\\ the second author's name\\ the second author's affiliation. More can be listed like this.}
% {$^*$ The titles and numbers of the foundations that support this article.}
%*************************************************************************************************************
\biaoti{A Functional Complexity Framework for the Analysis of Telecommunication Networks$^*$}%%%   Main Title of your paper  %%%
{\uppercase{Dzaferagic} Merim \cdd \uppercase{Kaminski}
Nicholas  \cdd \uppercase{McBride} Neal \cdd \uppercase{Macaluso} Irene \cdd \uppercase{Marchetti} Nicola}%%% The names of the authors  %%%
{Address\\
    Email: dzaferam@tcd.ie; kaminskn@tcd.ie; mcbridne@tcd.ie; macalusi@tcd.ie; marchetn@tcd.ie} %%% The address of the authors  %%%
{$^*$This material is based upon works supported by the Science Foundation Ireland under Grant No. 13/RC/2077
%{$^\diamond${\it This paper was recommended for publication byEditor . }}
}

%*************************************************************************************************************
%The submission date of your article. For example: \drd{Received: June 8, 2006}
%*************************************************************************************************************
\drd{DOI: }{Received: x x 20xx}{ / Revised: x x 20xx}

%*************************************************************************************************************
% The page header of the article.
% \dshm{Year}{Volume}{The capitalized RUNNING HEAD of your article with less than 48 letters}{The capitalized
% AUTHORS list with $\cdot$ separating different names or one can type "The name of the first author et al."
% if there are more than 4 authors.}
%*************************************************************************************************************

\dshm{20XX}{XX}{A Functional Complexity Framework for the Analysis of Telecommunication Networks}{\uppercase{Dzaferagic
Merim}, et al.}

%*************************************************************************************************************
% \dab{The abstract}{Keywords}
%*************************************************************************************************************
%-------------------------------------------------------------------------
\Abstract{\input{sections/abstract_section}}      % the abstract

\Keywords{Complex system science, functional complexity framework, telecommunication networks, structural complexity, frequency allocation. }        % the keywords

%\MRSubClass{05B05, 05B25, 20B25}      % MR(2000) Subject Classification

%\baselineskip 15pt

\section{Introduction}\label{sec:introduction}
\input{sections/introduction_section}
\section{Network Function Framework}\label{sec:framework}
\input{sections/framework_section}

\section{Complexity Model}\label{sec:complexityModel}
\input{sections/complexity_model_section_nicked}

\section{Analysis}\label{sec:Analysis}
\input{sections/analysisNew_section}

\section{Conclusion}\label{sec:conclulsion}
\input{sections/conclusion_section}

\input{sections/acronyms_section}

%\acknowledgements{\rm This material is based upon works supported by the Science Foundation Ireland under Grant No. 13/RC/2077}
%% Please thank the anonymous people who make contributions to this article. If you don't want it, please delete it.

%%Please make sure that your given name is abbreviated as the first capital letter, such as Zhang X T, Tami T,...

\end{document}

%% file: jsscN.tex
\makeatletter
\def\@oddfoot{\hfill}
\newcount\shumeicount
\def\setshumei#1#2#3{%
  \shumeicount=\count0
  \def\@oddhead{%
    \raise-5pt\hbox to0pt{\vrule width\hsize height 0pt depth 0.4pt\hss}\relax
    \ifnum \shumeicount=\count0
      \raise-7pt\hbox to0pt{\vrule width\hsize height 0pt depth 0.4pt\hss}\relax
      #1
    \else
      \ifodd\count0
        #2
      \else
        #3
       \fi
     \fi
  }%
}
\makeatother
\makeatletter
\def\@oddfoot{\hfill}
\newcount\shujiaocount
\def\setshujiao{%
  \shujiaocount=\count0
  \def\@oddfoot{%
      \ifodd\count0
      \else
      \fi
  }%
}
\makeatother
\def\biaoti#1#2#3#4{{
  \vspace*{0.3cm}
  \begin{flushleft} \Large\bf #1\end{flushleft}
  \vspace*{-0.2cm}
      \begin{flushleft}
      \bf #2
      \end{flushleft}
      \footnotetext{\hspace{-6mm} #3\\ #4}}}
\def\dshm#1#2#3#4
{\setshumei{J Syst Sci Complex (#1) #2:
{\thepage--\pageref{LastPage}}\hfill}
            {\hfill {\small #3}\hfill\hbox to0pt{\hss\thepage}}
            {\hbox to0pt{\thepage\hss}\hfill {\small #4}\hfill
            }
            \setshujiao}
\def\drd#1#2
{{\vskip 1cm\small \begin{flushleft}
 #1 \\
 #2\\
\copyright The Editorial Office of JSSC \&  Springer-Verlag Berlin
Heidelberg 2014
\end{flushleft}}}
\def\epsilon{\varepsilon}

%% file: sections/abstract_section.tex
% !TeX spellcheck = en_GB
The rapid evolution of network services demands new paradigms for studying and designing networks. \textcolor{lastUpdate}{We propose a framework to investigate the underlying mechanisms of wireless network functions.}
%In order to understand the underlying mechanisms that provide network functions, we propose a framework which enables the functional analysis of telecommunication networks. 
\textcolor{lastUpdate}{This framework isolates and analyses a network function as a complex system. We propose functional topologies to visualise and systematically study the relationships between system entities.} We also define a complexity metric $C_F$ (functional complexity) which quantifies the variety of structural patterns and roles of nodes in the topology. This complexity metric provides a wholly new approach to study the operation of telecommunication networks. We study the relationship between $C_F$ and graph structures by analysing graph theory metrics in order to recognize complex organisations. $C_F$ is equal to zero for both a full mesh topology and a disconnected topology. \textcolor{activecolor}{We show that complexity is high for a structure with shorter average path length and higher average clustering coefficient}. We make a connection between functional complexity, robustness and response to changes that may appear in the system configuration. We also make a connection between the implementation and the outcome of a network function which correlates the characteristics of the outcome with the complex relationships that underpin the functional structure. 

%% file: sections/introduction_section.tex
% !TeX spellcheck = en_GB
The transition of humanity into the Information Age has precipitated the need of new paradigms to comprehend and overcome a new set of challenges. Specifically, the telecommunications networks that underpin modern societies represent some of the largest scale construction and deployment efforts ever attempted by humanity, with renovations occurring nearly continuously over the course of decades. The result is networks that consist of numerous subsections, each of which following its own trajectory of development, commingled into a complex cacophony. Considering the high degree of heterogeneity and dense interplay of network elements in proposed 5G and \ac{IoT} systems, achieving holistic understanding of network operation is poised to become an even more challenging prospect in the near future. The focus of our paper is to provide the paradigms necessary for comprehension of the multi-faceted, intermingled, and vital foundation of new networks by introducing a metric that quantifies the organisational structure of network functions.

Every telecommunication network is designed to provide different services. Network functions are the building blocks of these services. Understanding the mechanisms that provide network functions implies understanding the function and thus the network itself. In order to analyse functional aspects of a telecommunication network we introduce a framework to map the network function into a functional topology. The functional topology enables a complex systems approach to analyse functions of telecommunications networks. We also introduce a metric which quantifies the complexity\footnote{With the term "complexity" we refer to a specific set of complex systems science quantities, related to the interactions between functional entities (rather than to the entities themselves).} of a particular implementation of a network function. Specifically, the proposed metric analyses the underlying communication between network entities which provide the network function.

\textcolor{lastUpdate}{Using complex systems science we can analyse network functions holistically. This interdisciplinary field draws attention from researchers in physics, mathematics, engineering and many others. As we plan future networks, the experience of these fields is useful to draw upon. } Papers like \cite{Liu2013,Bar-Yam2004,Yaeger2007,Tononi1994,Lopez-Ruiz1995,Balduzzi2008,Joshi2013} testify about the interdisciplinary nature of the complex systems analysis. The authors of \cite{Evans2002} discuss issues of different research areas. This discussion leads to the conclusion that different scientific fields face similar problems. An interdisciplinary approach, which implies borrowing solutions from other scientific fields, would save time and effort that researchers put into solving problems. 

Different sciences faced the problem of increasing complexity differently. In \cite{Lloyd2001} the author presents a categorization of different complexity metrics and emphasises the interdisciplinary applicability of these metrics. The authors of \cite{Wang2014,Ahn2010,Evans2009,Newman2003,Newman2006} focus on community detection which allows us to analyse the structural organisation of a complex system. In \cite{Wang2014} the authors investigate the underlying interactions between mobile phone users which determine the affiliation to a community. Similarly, in \cite{Newman2003,Newman2006} the authors focus on the modularity of the system. The authors of \cite{Ahn2010,Evans2009} focus on links rather than nodes, which enables the detection of overlapping communities. Additionally, their approach allows them to analyse the hierarchical structure of a complex system. In contrast, we focus on the structural features of the graph representation of a network function implementation, rather than on community detection.

The authors of \cite{Lanham2013,Bristow2014,Gonzalez2008} examine social networks and user behaviour from a complex systems science perspective. In \cite{Lanham2013}, the authors propose a data-to-model process which allows them to analyse complex social interactions. This approach enables the prediction of the developments of eventual disasters in the system, which makes it possible to prepare the disaster recovery scenarios. The authors of \cite{Bristow2014} propose an agent-based framework for modelling competitive and cooperative behaviour under conflict (i.e. \ac{CPR} Dilemma). The framework allows them to study how and why do we reach some outcomes, and to determine the conditions needed to achieve desirable outcomes in a complex system. In \cite{Gonzalez2008}, the authors analyse the human travel patterns based on the trajectory of 100,000 mobile phone users. The understanding of the mobility patterns allows them to predict the movement and therefore the influence on spreading viruses, urban planning, mobile network planning, etc. We propose a framework that treats the implementation of a network function as a complex system, which allows us to focus on the network function itself, rather than the impact of the unpredictability of human behaviour on the network. 

Authors of \cite{Wang2009,Onnela2007,Beigy2010,Macaluso2014,Macaluso2016} analyse telecommunications networks as complex systems. The approach in \cite{Wang2009,Onnela2007} involves analysis of complex phenomena in telecommunication networks (spreading patterns of mobile viruses and connection strengths between nodes in a social network) which are the result of complex user behaviour. The authors of \cite{Macaluso2014,Macaluso2016} analyse the complexity of outcomes of a self-organising frequency allocation algorithm. In \cite{Macaluso2016}, they analysed a relationship between robustness and complexity of the outcome. In contrast to the approach in \cite{Wang2009,Onnela2007,Macaluso2014,Macaluso2016,Beigy2010}, the work we propose here targets the network function itself, which means that we analyse the mechanisms that enable complex user behaviour and system outcome. We use the same frequency allocation algorithm proposed in \cite{Macaluso2016} to show an example of how to map a network function (i.e. frequency allocation) into a functional topology. 

The main contributions of this paper are:
\begin{itemize}
	\item We introduce a framework which enables the functional analysis of a telecommunication network;
	\item We provide several examples that show how to map a network function into a functional topology;
	\item We provide a new complexity metric that quantifies the organisational structure of a telecommunication network function.
%	\item We quantify the organisational structure of telecommunication functions
\end{itemize}

%% file: sections/framework_section.tex
% !TeX spellcheck = en_GB
%TODO FINID APPROPRIATE REFERENCE MAYBE SOMETHING ABOUT SOFTWARE DEFINED RADIO 
The approach to planning, configuration, management and optimization of network functions is changing and moving towards self-organisation. The traditional approach to most network functions involves the use of central control or optimization. However, the increasing heterogeneity of wireless technologies contributes to the rapid evolution, change, and growth of networks that makes the centralised approach unsustainable. If controlled in a self-organizing way, network functions such as handover, transmit power control, user allocation, data rate control and frequency allocation provide more flexibility and robustness in response to changes that may appear in the network \cite{kaminski2014social}. %TODO find appropriate references for diferent self organizing algorithms for network functions (hing Ryan thomas 2007 thesis he has a survay or something)

In order to specify and analyse the complexity of a network function, we introduce the \textit{functional framework}. Our framework represents an abstraction of a telecommunication network modelling its operation by capturing all elements, i.e. nodes and connections, necessary to perform a given function. Our framework includes \textit{functional topologies} which are graphs created based on the functional connectivity between system entities. In contrast to logical topologies which refer to data flows between network nodes, functional topologies depict the connectivity pattern of a network function, which implies a broader meaning to nodes and links. %nodes in order to analyse the complexity of the given function. 

A node in our topology represents a functional entity of a network node or any information source that is part of the given network function. The links indicate dependencies between nodes. The topology as a whole depicts the specific implementation of the network function. The topology allows us to apply our model to analyse the complexity of the implementation. The functional complexity quantifies the organisational structure by analysing the variety of relationships between system entities and roles that these entities have in the topology. 

As an instructive example, we focus on self-organization from a frequency allocation perspective. More precisely, we use the frequency allocation algorithm from \cite{Macaluso2016} to describe the process of mapping a network function into a topology and in the end to calculate the functional complexity.

The self-organising frequency allocation algorithm considers a cellular network as a two-dimensional cellular automaton. Each cell in the model represents a self-organizing wireless system. The algorithm works based on the local information that nodes gather from their neighbours. Briefly, every cell senses the given frequency channels, and allocates a channel with no interference. For more details about the algorithm the reader is referred to \cite{Macaluso2016}. 

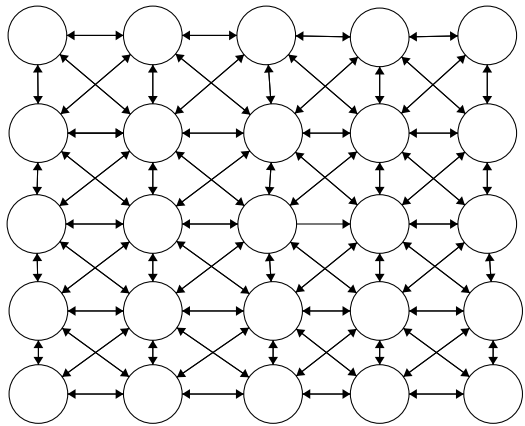
\begin{figure}
\begin{center}
		\begin{tikzpicture}[scale=0.13]
		\tikzstyle{every node}+=[inner sep=0pt]
		\draw [black] (26.8,-9.5) circle (3);
		\draw [black] (38.4,-9.5) circle (3);
		\draw [black] (50,-9.7) circle (3);
		\draw [black] (15,-9.5) circle (3);
		\draw [black] (61,-9.5) circle (3);
		\draw [black] (15.1,-19.5) circle (3);
		\draw [black] (14.9,-28.8) circle (3);
		\draw [black] (15.1,-37.7) circle (3);
		\draw [black] (15.1,-46.2) circle (3);
		\draw [black] (26.8,-19.5) circle (3);
		\draw [black] (39.1,-19.5) circle (3);
		\draw [black] (50,-19.5) circle (3);
		\draw [black] (61,-19.5) circle (3);
		\draw [black] (26.8,-28.8) circle (3);
		\draw [black] (38.5,-28.8) circle (3);
		\draw [black] (50,-28.8) circle (3);
		\draw [black] (61,-28.8) circle (3);
		\draw [black] (26.8,-37.7) circle (3);
		\draw [black] (39.1,-37.7) circle (3);
		\draw [black] (50,-37.7) circle (3);
		\draw [black] (26.8,-46.2) circle (3);
		\draw [black] (39.1,-46.2) circle (3);
		\draw [black] (50,-46.2) circle (3);
		\draw [black] (61.6,-46.2) circle (3);
		\draw [black] (61.6,-37.7) circle (3);
		\draw [black] (18,-9.5) -- (23.8,-9.5);
		\fill [black] (23.8,-9.5) -- (23,-9) -- (23,-10);
		\draw [black] (23.8,-9.5) -- (18,-9.5);
		\fill [black] (18,-9.5) -- (18.8,-10) -- (18.8,-9);
		\draw [black] (29.8,-9.5) -- (35.4,-9.5);
		\fill [black] (35.4,-9.5) -- (34.6,-9) -- (34.6,-10);
		\draw [black] (35.4,-9.5) -- (29.8,-9.5);
		\fill [black] (29.8,-9.5) -- (30.6,-10) -- (30.6,-9);
		\draw [black] (41.4,-9.55) -- (47,-9.65);
		\fill [black] (47,-9.65) -- (46.21,-9.13) -- (46.19,-10.13);
		\draw [black] (47,-9.65) -- (41.4,-9.55);
		\fill [black] (41.4,-9.55) -- (42.19,-10.07) -- (42.21,-9.07);
		\draw [black] (53,-9.65) -- (58,-9.55);
		\fill [black] (58,-9.55) -- (57.19,-9.07) -- (57.21,-10.07);
		\draw [black] (58,-9.55) -- (53,-9.65);
		\fill [black] (53,-9.65) -- (53.81,-10.13) -- (53.79,-9.13);
		\draw [black] (18.1,-19.5) -- (23.8,-19.5);
		\fill [black] (23.8,-19.5) -- (23,-19) -- (23,-20);
		\draw [black] (23.8,-19.5) -- (18.1,-19.5);
		\fill [black] (18.1,-19.5) -- (18.9,-20) -- (18.9,-19);
		\draw [black] (29.8,-19.5) -- (36.1,-19.5);
		\fill [black] (36.1,-19.5) -- (35.3,-19) -- (35.3,-20);
		\draw [black] (36.1,-19.5) -- (29.8,-19.5);
		\fill [black] (29.8,-19.5) -- (30.6,-20) -- (30.6,-19);
		\draw [black] (47,-19.5) -- (42.1,-19.5);
		\fill [black] (42.1,-19.5) -- (42.9,-20) -- (42.9,-19);
		\draw [black] (42.1,-19.5) -- (47,-19.5);
		\fill [black] (47,-19.5) -- (46.2,-19) -- (46.2,-20);
		\draw [black] (58,-19.5) -- (53,-19.5);
		\fill [black] (53,-19.5) -- (53.8,-20) -- (53.8,-19);
		\draw [black] (53,-19.5) -- (58,-19.5);
		\fill [black] (58,-19.5) -- (57.2,-19) -- (57.2,-20);
		\draw [black] (61,-22.5) -- (61,-25.8);
		\fill [black] (61,-25.8) -- (61.5,-25) -- (60.5,-25);
		\draw [black] (61,-25.8) -- (61,-22.5);
		\fill [black] (61,-22.5) -- (60.5,-23.3) -- (61.5,-23.3);
		\draw [black] (61,-16.5) -- (61,-12.5);
		\fill [black] (61,-12.5) -- (60.5,-13.3) -- (61.5,-13.3);
		\draw [black] (61,-12.5) -- (61,-16.5);
		\fill [black] (61,-16.5) -- (61.5,-15.7) -- (60.5,-15.7);
		\draw [black] (52.22,-17.48) -- (58.78,-11.52);
		\fill [black] (58.78,-11.52) -- (57.85,-11.69) -- (58.52,-12.43);
		\draw [black] (58.78,-11.52) -- (52.22,-17.48);
		\fill [black] (52.22,-17.48) -- (53.15,-17.31) -- (52.48,-16.57);
		\draw [black] (58.76,-17.5) -- (52.24,-11.7);
		\fill [black] (52.24,-11.7) -- (52.5,-12.6) -- (53.17,-11.85);
		\draw [black] (52.24,-11.7) -- (58.76,-17.5);
		\fill [black] (58.76,-17.5) -- (58.5,-16.6) -- (57.83,-17.35);
		\draw [black] (50,-12.7) -- (50,-16.5);
		\fill [black] (50,-16.5) -- (50.5,-15.7) -- (49.5,-15.7);
		\draw [black] (50,-16.5) -- (50,-12.7);
		\fill [black] (50,-12.7) -- (49.5,-13.5) -- (50.5,-13.5);
		\draw [black] (52.29,-21.44) -- (58.71,-26.86);
		\fill [black] (58.71,-26.86) -- (58.42,-25.96) -- (57.78,-26.73);
		\draw [black] (58.71,-26.86) -- (52.29,-21.44);
		\fill [black] (52.29,-21.44) -- (52.58,-22.34) -- (53.22,-21.57);
		\draw [black] (58,-28.8) -- (53,-28.8);
		\fill [black] (53,-28.8) -- (53.8,-29.3) -- (53.8,-28.3);
		\draw [black] (53,-28.8) -- (58,-28.8);
		\fill [black] (58,-28.8) -- (57.2,-28.3) -- (57.2,-29.3);
		\draw [black] (50,-25.8) -- (50,-22.5);
		\fill [black] (50,-22.5) -- (49.5,-23.3) -- (50.5,-23.3);
		\draw [black] (50,-22.5) -- (50,-25.8);
		\fill [black] (50,-25.8) -- (50.5,-25) -- (49.5,-25);
		\draw [black] (52.38,-30.63) -- (59.22,-35.87);
		\fill [black] (59.22,-35.87) -- (58.89,-34.99) -- (58.28,-35.78);
		\draw [black] (59.22,-35.87) -- (52.38,-30.63);
		\fill [black] (52.38,-30.63) -- (52.71,-31.51) -- (53.32,-30.72);
		\draw [black] (61.2,-31.79) -- (61.4,-34.71);
		\fill [black] (61.4,-34.71) -- (61.84,-33.87) -- (60.85,-33.94);
		\draw [black] (61.4,-34.71) -- (61.2,-31.79);
		\fill [black] (61.2,-31.79) -- (60.76,-32.63) -- (61.75,-32.56);
		\draw [black] (50,-31.8) -- (50,-34.7);
		\fill [black] (50,-34.7) -- (50.5,-33.9) -- (49.5,-33.9);
		\draw [black] (50,-34.7) -- (50,-31.8);
		\fill [black] (50,-31.8) -- (49.5,-32.6) -- (50.5,-32.6);
		\draw [black] (53,-37.7) -- (58.6,-37.7);
		\fill [black] (58.6,-37.7) -- (57.8,-37.2) -- (57.8,-38.2);
		\draw [black] (58.6,-37.7) -- (53,-37.7);
		\fill [black] (53,-37.7) -- (53.8,-38.2) -- (53.8,-37.2);
		\draw [black] (52.42,-39.47) -- (59.18,-44.43);
		\fill [black] (59.18,-44.43) -- (58.83,-43.55) -- (58.24,-44.36);
		\draw [black] (59.18,-44.43) -- (52.42,-39.47);
		\fill [black] (52.42,-39.47) -- (52.77,-40.35) -- (53.36,-39.54);
		\draw [black] (59.18,-39.47) -- (52.42,-44.43);
		\fill [black] (52.42,-44.43) -- (53.36,-44.36) -- (52.77,-43.55);
		\draw [black] (52.42,-44.43) -- (59.18,-39.47);
		\fill [black] (59.18,-39.47) -- (58.24,-39.54) -- (58.83,-40.35);
		\draw [black] (61.6,-40.7) -- (61.6,-43.2);
		\fill [black] (61.6,-43.2) -- (62.1,-42.4) -- (61.1,-42.4);
		\draw [black] (61.6,-43.2) -- (61.6,-40.7);
		\fill [black] (61.6,-40.7) -- (61.1,-41.5) -- (62.1,-41.5);
		\draw [black] (58.67,-30.69) -- (52.33,-35.81);
		\fill [black] (52.33,-35.81) -- (53.27,-35.7) -- (52.64,-34.92);
		\draw [black] (52.33,-35.81) -- (58.67,-30.69);
		\fill [black] (58.67,-30.69) -- (57.73,-30.8) -- (58.36,-31.58);
		\draw [black] (52.29,-26.86) -- (58.71,-21.44);
		\fill [black] (58.71,-21.44) -- (57.78,-21.57) -- (58.42,-22.34);
		\draw [black] (58.71,-21.44) -- (52.29,-26.86);
		\fill [black] (52.29,-26.86) -- (53.22,-26.73) -- (52.58,-25.96);
		\draw [black] (50,-40.7) -- (50,-43.2);
		\fill [black] (50,-43.2) -- (50.5,-42.4) -- (49.5,-42.4);
		\draw [black] (50,-43.2) -- (50,-40.7);
		\fill [black] (50,-40.7) -- (49.5,-41.5) -- (50.5,-41.5);
		\draw [black] (41.33,-17.49) -- (47.77,-11.71);
		\fill [black] (47.77,-11.71) -- (46.84,-11.87) -- (47.51,-12.61);
		\draw [black] (47.73,-17.54) -- (40.67,-11.46);
		\fill [black] (40.67,-11.46) -- (40.95,-12.36) -- (41.6,-11.6);
		\draw [black] (40.67,-11.46) -- (47.73,-17.54);
		\fill [black] (47.73,-17.54) -- (47.45,-16.64) -- (46.8,-17.4);
		\draw [black] (47.77,-11.71) -- (41.33,-17.49);
		\fill [black] (41.33,-17.49) -- (42.26,-17.33) -- (41.59,-16.59);
		\draw [black] (38.89,-16.51) -- (38.61,-12.49);
		\fill [black] (38.61,-12.49) -- (38.17,-13.33) -- (39.16,-13.26);
		\draw [black] (38.61,-12.49) -- (38.89,-16.51);
		\fill [black] (38.89,-16.51) -- (39.33,-15.67) -- (38.34,-15.74);
		\draw [black] (38.91,-22.49) -- (38.69,-25.81);
		\fill [black] (38.69,-25.81) -- (39.24,-25.04) -- (38.25,-24.98);
		\draw [black] (38.69,-25.81) -- (38.91,-22.49);
		\fill [black] (38.91,-22.49) -- (38.36,-23.26) -- (39.35,-23.32);
		\draw [black] (40.83,-26.91) -- (47.67,-21.39);
		\fill [black] (47.67,-21.39) -- (46.73,-21.5) -- (47.36,-22.28);
		\draw [black] (47.67,-21.39) -- (40.83,-26.91);
		\fill [black] (40.83,-26.91) -- (41.77,-26.8) -- (41.14,-26.02);
		\draw [black] (41.5,-28.8) -- (47,-28.8);
		\fill [black] (47,-28.8) -- (46.2,-28.3) -- (46.2,-29.3);
		\draw [black] (47.72,-26.85) -- (41.38,-21.45);
		\fill [black] (41.38,-21.45) -- (41.67,-22.35) -- (42.32,-21.59);
		\draw [black] (41.38,-21.45) -- (47.72,-26.85);
		\fill [black] (47.72,-26.85) -- (47.43,-25.95) -- (46.78,-26.71);
		\draw [black] (47.68,-30.7) -- (41.42,-35.8);
		\fill [black] (41.42,-35.8) -- (42.36,-35.68) -- (41.73,-34.91);
		\draw [black] (41.42,-35.8) -- (47.68,-30.7);
		\fill [black] (47.68,-30.7) -- (46.74,-30.82) -- (47.37,-31.59);
		\draw [black] (47.63,-35.86) -- (40.87,-30.64);
		\fill [black] (40.87,-30.64) -- (41.2,-31.52) -- (41.81,-30.73);
		\draw [black] (40.87,-30.64) -- (47.63,-35.86);
		\fill [black] (47.63,-35.86) -- (47.3,-34.98) -- (46.69,-35.77);
		\draw [black] (38.9,-34.71) -- (38.7,-31.79);
		\fill [black] (38.7,-31.79) -- (38.26,-32.63) -- (39.25,-32.56);
		\draw [black] (38.7,-31.79) -- (38.9,-34.71);
		\fill [black] (38.9,-34.71) -- (39.34,-33.87) -- (38.35,-33.94);
		\draw [black] (39.1,-40.7) -- (39.1,-43.2);
		\fill [black] (39.1,-43.2) -- (39.6,-42.4) -- (38.6,-42.4);
		\draw [black] (39.1,-43.2) -- (39.1,-40.7);
		\fill [black] (39.1,-40.7) -- (38.6,-41.5) -- (39.6,-41.5);
		\draw [black] (42.1,-37.7) -- (47,-37.7);
		\fill [black] (47,-37.7) -- (46.2,-37.2) -- (46.2,-38.2);
		\draw [black] (47,-37.7) -- (42.1,-37.7);
		\fill [black] (42.1,-37.7) -- (42.9,-38.2) -- (42.9,-37.2);
		\draw [black] (41.47,-39.54) -- (47.63,-44.36);
		\fill [black] (47.63,-44.36) -- (47.31,-43.47) -- (46.7,-44.26);
		\draw [black] (47.63,-44.36) -- (41.47,-39.54);
		\fill [black] (41.47,-39.54) -- (41.79,-40.43) -- (42.4,-39.64);
		\draw [black] (41.47,-44.36) -- (47.63,-39.54);
		\fill [black] (47.63,-39.54) -- (46.7,-39.64) -- (47.31,-40.43);
		\draw [black] (47.63,-39.54) -- (41.47,-44.36);
		\fill [black] (41.47,-44.36) -- (42.4,-44.26) -- (41.79,-43.47);
		\draw [black] (42.1,-46.2) -- (47,-46.2);
		\fill [black] (47,-46.2) -- (46.2,-45.7) -- (46.2,-46.7);
		\draw [black] (47,-46.2) -- (42.1,-46.2);
		\fill [black] (42.1,-46.2) -- (42.9,-46.7) -- (42.9,-45.7);
		\draw [black] (29.13,-11.39) -- (36.77,-17.61);
		\fill [black] (36.77,-17.61) -- (36.47,-16.71) -- (35.84,-17.49);
		\draw [black] (36.77,-17.61) -- (29.13,-11.39);
		\fill [black] (29.13,-11.39) -- (29.43,-12.29) -- (30.06,-11.51);
		\draw [black] (26.8,-12.5) -- (26.8,-16.5);
		\fill [black] (26.8,-16.5) -- (27.3,-15.7) -- (26.3,-15.7);
		\draw [black] (26.8,-16.5) -- (26.8,-12.5);
		\fill [black] (26.8,-12.5) -- (26.3,-13.3) -- (27.3,-13.3);
		\draw [black] (29.07,-17.54) -- (36.13,-11.46);
		\fill [black] (36.13,-11.46) -- (35.2,-11.6) -- (35.85,-12.36);
		\draw [black] (36.13,-11.46) -- (29.07,-17.54);
		\fill [black] (29.07,-17.54) -- (30,-17.4) -- (29.35,-16.64);
		\draw [black] (29.15,-21.37) -- (36.15,-26.93);
		\fill [black] (36.15,-26.93) -- (35.84,-26.04) -- (35.21,-26.83);
		\draw [black] (36.15,-26.93) -- (29.15,-21.37);
		\fill [black] (29.15,-21.37) -- (29.46,-22.26) -- (30.09,-21.47);
		\draw [black] (26.8,-22.5) -- (26.8,-25.8);
		\fill [black] (26.8,-25.8) -- (27.3,-25) -- (26.3,-25);
		\draw [black] (26.8,-25.8) -- (26.8,-22.5);
		\fill [black] (26.8,-22.5) -- (26.3,-23.3) -- (27.3,-23.3);
		\draw [black] (26.8,-31.8) -- (26.8,-34.7);
		\fill [black] (26.8,-34.7) -- (27.3,-33.9) -- (26.3,-33.9);
		\draw [black] (26.8,-34.7) -- (26.8,-31.8);
		\fill [black] (26.8,-31.8) -- (26.3,-32.6) -- (27.3,-32.6);
		\draw [black] (26.8,-40.7) -- (26.8,-43.2);
		\fill [black] (26.8,-43.2) -- (27.3,-42.4) -- (26.3,-42.4);
		\draw [black] (26.8,-43.2) -- (26.8,-40.7);
		\fill [black] (26.8,-40.7) -- (26.3,-41.5) -- (27.3,-41.5);
		\draw [black] (29.8,-37.7) -- (36.1,-37.7);
		\fill [black] (36.1,-37.7) -- (35.3,-37.2) -- (35.3,-38.2);
		\draw [black] (36.1,-37.7) -- (29.8,-37.7);
		\fill [black] (29.8,-37.7) -- (30.6,-38.2) -- (30.6,-37.2);
		\draw [black] (29.8,-28.8) -- (35.5,-28.8);
		\fill [black] (35.5,-28.8) -- (34.7,-28.3) -- (34.7,-29.3);
		\draw [black] (35.5,-28.8) -- (29.8,-28.8);
		\fill [black] (29.8,-28.8) -- (30.6,-29.3) -- (30.6,-28.3);
		\draw [black] (29.19,-26.99) -- (36.71,-21.31);
		\fill [black] (36.71,-21.31) -- (35.77,-21.39) -- (36.37,-22.19);
		\draw [black] (36.71,-21.31) -- (29.19,-26.99);
		\fill [black] (29.19,-26.99) -- (30.13,-26.91) -- (29.53,-26.11);
		\draw [black] (29.23,-30.56) -- (36.67,-35.94);
		\fill [black] (36.67,-35.94) -- (36.31,-35.07) -- (35.73,-35.88);
		\draw [black] (36.67,-35.94) -- (29.23,-30.56);
		\fill [black] (29.23,-30.56) -- (29.59,-31.43) -- (30.17,-30.62);
		\draw [black] (29.19,-35.88) -- (36.11,-30.62);
		\fill [black] (36.11,-30.62) -- (35.17,-30.7) -- (35.78,-31.5);
		\draw [black] (36.11,-30.62) -- (29.19,-35.88);
		\fill [black] (29.19,-35.88) -- (30.13,-35.8) -- (29.52,-35);
		\draw [black] (29.27,-39.41) -- (36.63,-44.49);
		\fill [black] (36.63,-44.49) -- (36.26,-43.63) -- (35.69,-44.45);
		\draw [black] (36.63,-44.49) -- (29.27,-39.41);
		\fill [black] (29.27,-39.41) -- (29.64,-40.27) -- (30.21,-39.45);
		\draw [black] (36.1,-46.2) -- (29.8,-46.2);
		\fill [black] (29.8,-46.2) -- (30.6,-46.7) -- (30.6,-45.7);
		\draw [black] (29.8,-46.2) -- (36.1,-46.2);
		\fill [black] (36.1,-46.2) -- (35.3,-45.7) -- (35.3,-46.7);
		\draw [black] (29.27,-44.49) -- (36.63,-39.41);
		\fill [black] (36.63,-39.41) -- (35.69,-39.45) -- (36.26,-40.27);
		\draw [black] (36.63,-39.41) -- (29.27,-44.49);
		\fill [black] (29.27,-44.49) -- (30.21,-44.45) -- (29.64,-43.63);
		\draw [black] (15.03,-12.5) -- (15.07,-16.5);
		\fill [black] (15.07,-16.5) -- (15.56,-15.7) -- (14.56,-15.71);
		\draw [black] (15.07,-16.5) -- (15.03,-12.5);
		\fill [black] (15.03,-12.5) -- (14.54,-13.3) -- (15.54,-13.29);
		\draw [black] (15.04,-22.5) -- (14.96,-25.8);
		\fill [black] (14.96,-25.8) -- (15.48,-25.01) -- (14.48,-24.99);
		\draw [black] (14.96,-25.8) -- (15.04,-22.5);
		\fill [black] (15.04,-22.5) -- (14.52,-23.29) -- (15.52,-23.31);
		\draw [black] (14.97,-31.8) -- (15.03,-34.7);
		\fill [black] (15.03,-34.7) -- (15.51,-33.89) -- (14.51,-33.91);
		\draw [black] (15.03,-34.7) -- (14.97,-31.8);
		\fill [black] (14.97,-31.8) -- (14.49,-32.61) -- (15.49,-32.59);
		\draw [black] (15.1,-40.7) -- (15.1,-43.2);
		\fill [black] (15.1,-43.2) -- (15.6,-42.4) -- (14.6,-42.4);
		\draw [black] (15.1,-43.2) -- (15.1,-40.7);
		\fill [black] (15.1,-40.7) -- (14.6,-41.5) -- (15.6,-41.5);
		\draw [black] (18.1,-46.2) -- (23.8,-46.2);
		\fill [black] (23.8,-46.2) -- (23,-45.7) -- (23,-46.7);
		\draw [black] (23.8,-46.2) -- (18.1,-46.2);
		\fill [black] (18.1,-46.2) -- (18.9,-46.7) -- (18.9,-45.7);
		\draw [black] (18.1,-37.7) -- (23.8,-37.7);
		\fill [black] (23.8,-37.7) -- (23,-37.2) -- (23,-38.2);
		\draw [black] (23.8,-37.7) -- (18.1,-37.7);
		\fill [black] (18.1,-37.7) -- (18.9,-38.2) -- (18.9,-37.2);
		\draw [black] (17.9,-28.8) -- (23.8,-28.8);
		\fill [black] (23.8,-28.8) -- (23,-28.3) -- (23,-29.3);
		\draw [black] (23.8,-28.8) -- (17.9,-28.8);
		\fill [black] (17.9,-28.8) -- (18.7,-29.3) -- (18.7,-28.3);
		\draw [black] (18.1,-19.5) -- (23.8,-19.5);
		\fill [black] (23.8,-19.5) -- (23,-19) -- (23,-20);
		\draw [black] (24.52,-11.45) -- (17.38,-17.55);
		\fill [black] (17.38,-17.55) -- (18.31,-17.41) -- (17.66,-16.65);
		\draw [black] (17.29,-11.44) -- (24.51,-17.56);
		\fill [black] (24.51,-17.56) -- (24.22,-16.66) -- (23.58,-17.42);
		\draw [black] (17.38,-17.55) -- (24.52,-11.45);
		\fill [black] (24.52,-11.45) -- (23.59,-11.59) -- (24.24,-12.35);
		\draw [black] (24.51,-17.56) -- (17.29,-11.44);
		\fill [black] (17.29,-11.44) -- (17.58,-12.34) -- (18.22,-11.58);
		\draw [black] (17.45,-21.37) -- (24.45,-26.93);
		\fill [black] (24.45,-26.93) -- (24.14,-26.04) -- (23.51,-26.83);
		\draw [black] (17.3,-30.6) -- (24.4,-35.9);
		\fill [black] (24.4,-35.9) -- (24.06,-35.02) -- (23.46,-35.82);
		\draw [black] (17.53,-39.46) -- (24.37,-44.44);
		\fill [black] (24.37,-44.44) -- (24.02,-43.56) -- (23.43,-44.37);
		\draw [black] (24.37,-44.44) -- (17.53,-39.46);
		\fill [black] (17.53,-39.46) -- (17.88,-40.34) -- (18.47,-39.53);
		\draw [black] (24.4,-35.9) -- (17.3,-30.6);
		\fill [black] (17.3,-30.6) -- (17.64,-31.48) -- (18.24,-30.68);
		\draw [black] (24.45,-26.93) -- (17.45,-21.37);
		\fill [black] (17.45,-21.37) -- (17.76,-22.26) -- (18.39,-21.47);
		\draw [black] (24.44,-21.35) -- (17.26,-26.95);
		\fill [black] (17.26,-26.95) -- (18.2,-26.85) -- (17.59,-26.07);
		\draw [black] (17.26,-26.95) -- (24.44,-21.35);
		\fill [black] (24.44,-21.35) -- (23.5,-21.45) -- (24.11,-22.23);
		\draw [black] (24.41,-30.62) -- (17.49,-35.88);
		\fill [black] (17.49,-35.88) -- (18.43,-35.8) -- (17.82,-35);
		\draw [black] (17.49,-35.88) -- (24.41,-30.62);
		\fill [black] (24.41,-30.62) -- (23.47,-30.7) -- (24.08,-31.5);
		\draw [black] (24.37,-39.46) -- (17.53,-44.44);
		\fill [black] (17.53,-44.44) -- (18.47,-44.37) -- (17.88,-43.56);
		\draw [black] (17.53,-44.44) -- (24.37,-39.46);
		\fill [black] (24.37,-39.46) -- (23.43,-39.53) -- (24.02,-40.34);
		\draw [black] (53,-46.2) -- (58.6,-46.2);
		\fill [black] (58.6,-46.2) -- (57.8,-45.7) -- (57.8,-46.7);
		\draw [black] (58.6,-46.2) -- (53,-46.2);
		\fill [black] (53,-46.2) -- (53.8,-46.7) -- (53.8,-45.7);
		\end{tikzpicture}\vskip3mm
	\caption{The physical topology according to the Moore neighbourhood from cellular automata.}
	\label{fig:physicalTopology}
	
\end{center}
\end{figure}

We focus on two frequency allocation algorithms (self-organizing algorithm and random frequency assignment) from \cite{Macaluso2016}. Here we analyse the implementation rather than the outcome of the function. Each autonomous network is modelled as a node in a lattice which means that the physical topology is the same for both implementations. The physical topology according to the Moore neighbourhood is shown in Figure \ref{fig:physicalTopology}. Different implementations of the network function are mapped into different functional topologies. 

By analysing the physical topology shown in Figure \ref{fig:physicalTopology}, we recognize a motif that represents the Moore neighbourhood. The motif consists of nine neighbouring cells, as shown in Figure \ref{fig:funcTopologySelfOrganising}. The motif depicts the local connections between neighbouring cells. The entire network is simply a repetition of that motif. The functional topology of an implementation of a network function is built upon the motif (Figure \ref{fig:templatesOfFuncTopologies}).

We start with the random frequency allocation algorithm. It assumes that every cell assigns the frequency completely randomly from the set of available frequency channels. In order to create the functional topology let us imagine a virtual decision maker entity that is moving from one cell to another. At every cell the decision maker entity has no information about the allocated frequencies of other cells, which means that there are no functional connections between any two nodes. The result is a functional topology represented with a non-connected graph (Figure \ref{fig:functionalTopologyRandomApproach}).

\begin{figure}
\centering \subfigure[\textcolor{lastUpdate}{The functional topology that represents the distributed self-organising frequency allocation algorithm; the Moore neighbourhood motif.}]{
	\begin{tikzpicture}[scale=0.13]
	\tikzstyle{every node}+=[inner sep=0pt]
	\draw [black] (26.8,-9.5) circle (3);
	\draw [black] (38.4,-9.5) circle (3);
	\draw [black] (15,-9.5) circle (3);
	\draw [black] (15.1,-19.5) circle (3);
	\draw [black] (14.9,-28.8) circle (3);
	\draw [black] (26.8,-19.5) circle (3);
	\draw [black] (38.5,-19.5) circle (3);
	\draw [black] (26.8,-28.8) circle (3);
	\draw [black] (38.5,-28.8) circle (3);
	\draw [black] (18,-9.5) -- (23.8,-9.5);
	\fill [black] (23.8,-9.5) -- (23,-9) -- (23,-10);
	\draw [black] (23.8,-9.5) -- (18,-9.5);
	\fill [black] (18,-9.5) -- (18.8,-10) -- (18.8,-9);
	\draw [black] (29.8,-9.5) -- (35.4,-9.5);
	\fill [black] (35.4,-9.5) -- (34.6,-9) -- (34.6,-10);
	\draw [black] (35.4,-9.5) -- (29.8,-9.5);
	\fill [black] (29.8,-9.5) -- (30.6,-10) -- (30.6,-9);
	\draw [black] (18.1,-19.5) -- (23.8,-19.5);
	\fill [black] (23.8,-19.5) -- (23,-19) -- (23,-20);
	\draw [black] (23.8,-19.5) -- (18.1,-19.5);
	\fill [black] (18.1,-19.5) -- (18.9,-20) -- (18.9,-19);
	\draw [black] (29.8,-19.5) -- (35.5,-19.5);
	\fill [black] (35.5,-19.5) -- (34.7,-19) -- (34.7,-20);
	\draw [black] (35.5,-19.5) -- (29.8,-19.5);
	\fill [black] (29.8,-19.5) -- (30.6,-20) -- (30.6,-19);
	\draw [black] (38.47,-16.5) -- (38.43,-12.5);
	\fill [black] (38.43,-12.5) -- (37.94,-13.3) -- (38.94,-13.29);
	\draw [black] (38.43,-12.5) -- (38.47,-16.5);
	\fill [black] (38.47,-16.5) -- (38.96,-15.7) -- (37.96,-15.71);
	\draw [black] (29.08,-11.45) -- (36.22,-17.55);
	\fill [black] (36.22,-17.55) -- (35.94,-16.65) -- (35.29,-17.41);
	\draw [black] (36.22,-17.55) -- (29.08,-11.45);
	\fill [black] (29.08,-11.45) -- (29.36,-12.35) -- (30.01,-11.59);
	\draw [black] (26.8,-12.5) -- (26.8,-16.5);
	\fill [black] (26.8,-16.5) -- (27.3,-15.7) -- (26.3,-15.7);
	\draw [black] (26.8,-16.5) -- (26.8,-12.5);
	\fill [black] (26.8,-12.5) -- (26.3,-13.3) -- (27.3,-13.3);
	\draw [black] (29.07,-17.54) -- (36.13,-11.46);
	\fill [black] (36.13,-11.46) -- (35.2,-11.6) -- (35.85,-12.36);
	\draw [black] (36.13,-11.46) -- (29.07,-17.54);
	\fill [black] (29.07,-17.54) -- (30,-17.4) -- (29.35,-16.64);
	\draw [black] (26.8,-22.5) -- (26.8,-25.8);
	\fill [black] (26.8,-25.8) -- (27.3,-25) -- (26.3,-25);
	\draw [black] (26.8,-25.8) -- (26.8,-22.5);
	\fill [black] (26.8,-22.5) -- (26.3,-23.3) -- (27.3,-23.3);
	\draw [black] (29.15,-26.93) -- (36.15,-21.37);
	\fill [black] (36.15,-21.37) -- (35.21,-21.47) -- (35.84,-22.26);
	\draw [black] (36.15,-21.37) -- (29.15,-26.93);
	\fill [black] (29.15,-26.93) -- (30.09,-26.83) -- (29.46,-26.04);
	\draw [black] (15.03,-12.5) -- (15.07,-16.5);
	\fill [black] (15.07,-16.5) -- (15.56,-15.7) -- (14.56,-15.71);
	\draw [black] (15.07,-16.5) -- (15.03,-12.5);
	\fill [black] (15.03,-12.5) -- (14.54,-13.3) -- (15.54,-13.29);
	\draw [black] (15.04,-22.5) -- (14.96,-25.8);
	\fill [black] (14.96,-25.8) -- (15.48,-25.01) -- (14.48,-24.99);
	\draw [black] (14.96,-25.8) -- (15.04,-22.5);
	\fill [black] (15.04,-22.5) -- (14.52,-23.29) -- (15.52,-23.31);
	\draw [black] (17.9,-28.8) -- (23.8,-28.8);
	\fill [black] (23.8,-28.8) -- (23,-28.3) -- (23,-29.3);
	\draw [black] (23.8,-28.8) -- (17.9,-28.8);
	\fill [black] (17.9,-28.8) -- (18.7,-29.3) -- (18.7,-28.3);
	\draw [black] (18.1,-19.5) -- (23.8,-19.5);
	\fill [black] (23.8,-19.5) -- (23,-19) -- (23,-20);
	\draw [black] (24.52,-11.45) -- (17.38,-17.55);
	\fill [black] (17.38,-17.55) -- (18.31,-17.41) -- (17.66,-16.65);
	\draw [black] (17.29,-11.44) -- (24.51,-17.56);
	\fill [black] (24.51,-17.56) -- (24.22,-16.66) -- (23.58,-17.42);
	\draw [black] (17.38,-17.55) -- (24.52,-11.45);
	\fill [black] (24.52,-11.45) -- (23.59,-11.59) -- (24.24,-12.35);
	\draw [black] (24.51,-17.56) -- (17.29,-11.44);
	\fill [black] (17.29,-11.44) -- (17.58,-12.34) -- (18.22,-11.58);
	\draw [black] (17.45,-21.37) -- (24.45,-26.93);
	\fill [black] (24.45,-26.93) -- (24.14,-26.04) -- (23.51,-26.83);
	\draw [black] (24.45,-26.93) -- (17.45,-21.37);
	\fill [black] (17.45,-21.37) -- (17.76,-22.26) -- (18.39,-21.47);
	\draw [black] (24.44,-21.35) -- (17.26,-26.95);
	\fill [black] (17.26,-26.95) -- (18.2,-26.85) -- (17.59,-26.07);
	\draw [black] (17.26,-26.95) -- (24.44,-21.35);
	\fill [black] (24.44,-21.35) -- (23.5,-21.45) -- (24.11,-22.23);
	\draw [black] (29.15,-21.37) -- (36.15,-26.93);
	\fill [black] (36.15,-26.93) -- (35.84,-26.04) -- (35.21,-26.83);
	\draw [black] (36.15,-26.93) -- (29.15,-21.37);
	\fill [black] (29.15,-21.37) -- (29.46,-22.26) -- (30.09,-21.47);
	\draw [black] (38.5,-22.5) -- (38.5,-25.8);
	\fill [black] (38.5,-25.8) -- (39,-25) -- (38,-25);
	\draw [black] (38.5,-25.8) -- (38.5,-22.5);
	\fill [black] (38.5,-22.5) -- (38,-23.3) -- (39,-23.3);
	\draw [black] (35.5,-28.8) -- (29.8,-28.8);
	\fill [black] (29.8,-28.8) -- (30.6,-29.3) -- (30.6,-28.3);
	\draw [black] (29.8,-28.8) -- (35.5,-28.8);
	\fill [black] (35.5,-28.8) -- (34.7,-28.3) -- (34.7,-29.3);
	\end{tikzpicture}
	\label{fig:funcTopologySelfOrganising} } 
\hskip9mm \subfigure[\textcolor{lastUpdate}{The functional topology that represents the random frequency allocation approach. Independent network functions have no interconnections}]{
				\begin{tikzpicture}[scale=0.13]
				\tikzstyle{every node}+=[inner sep=0pt]
				\draw [black] (26.8,-9.5) circle (3);
				\draw [black] (38.4,-9.5) circle (3);
				\draw [black] (15,-9.5) circle (3);
				\draw [black] (15.1,-19.5) circle (3);
				\draw [black] (14.9,-28.8) circle (3);
				\draw [black] (26.8,-19.5) circle (3);
				\draw [black] (38.5,-19.5) circle (3);
				\draw [black] (26.8,-28.8) circle (3);
				\draw [black] (38.5,-28.8) circle (3);
				\end{tikzpicture}
	\label{fig:functionalTopologyRandomApproach} }

\caption{}
\label{fig:templatesOfFuncTopologies}
%\centerline {\small \parbox[t]{1.7cm}{\bf Figure 2}
%	\parbox[t]{10.5cm}
%	{The estimated function $\widehat{g}(\cdot)$ (Figure(a))
%		and the corresponding scatter plot (Figure(b)) of the residuals,
%		where $Y = medv-mean(medv)$, $Z$ = (\textit{rm}, $\log$(\textit{tax}), \textit{ptratio},
%		$\log$(\textit{lstat}))$^\tau$.}\label{fig:templatesOfFuncTopologies}}

\end{figure}

To examine the functional topology of the self-organising frequency allocation algorithm we use the same approach presented in the previous example. We imagine a virtual decision maker entity which is responsible for the frequency allocation of every node. In order to determine the functional connections of a node we analyse the interactions to other nodes in the process of decision making at our target node. As the self-organising algorithm assumes only the knowledge about allocated frequencies at the neighbouring cells, the functional connections exist only between physical neighbours. This results in a functional topology that is equivalent to the Moore neighbourhood motif of the physical topology (Figure \ref{fig:funcTopologySelfOrganising}).

\textcolor{lastUpdate}{In general network functions are not as simple as the above example. Therefore, to analyse the organisational structure of a network function we apply a multi-scale approach.} The functional topology itself is built upon the local relationships between network cells according to the specified frequency allocation algorithm. The local interactions represent the lowest scale size. Analysis of higher scales is enabled by a multi-hop interaction examination. The reachability among nodes represents the interactions between them. By restricting the reachability to a certain number of hops along the topology we may examine the scale size of interest. 

In \cite{Bar-Yam2004} the authors discussed the relation between the scale size and the complexity, and they emphasize the importance of the scale size noting that the same system can show high complexity on one scale and low complexity on the other. The functional connectivity of each node represents the lowest scale in the functional framework and describes the interactions in the process of making a decision on this particular node. From the functional point of view the lowest scale size models a decision that is made by one node (in our case the frequency allocation of one cell). In order to apply a multi-scale approach we include higher scales in our analysis as follows. As every node represents a functional part, a subtask or an informational source, every group of nodes represents a group of subtasks that are part of the same function. The 2-hop scale size considers the node, its neighbouring cells and neighbours of its neighbours. On this scale size the topology includes the interactions of nodes that are inside the radius of two hops. In our examples this scale size represents the frequency allocation of a two hop neighbourhood and the interactions between nodes in the process of making a decision among a group of nodes. Note that the reachability of nodes increases with the scale size. In this case every node can reach any other node that is two or less than two hops away. The highest scale size implies a full mesh topology, where every node can reach any other node in the topology. This approach enables a multi-scale functional analysis which is presented in the results section. 

%TODO
%2. ???? Stupid node <-> intelligent network (stupid scale size)

%DONE
%We start with the centralised approach. The centralised approach assumes knowledge about frequency channels allocated at all cells in every moment. In order to create the functional topology let us imagine a virtual decision maker entity that is moving from one cell to another. On every node the virtual decision maker entity has knowledge about allocated frequencies at all other nodes, which means that every node has a direct functional link to every other node. This results in a full mesh graph that represents the functional topology of the centralised approach to assign frequencies (Figure \ref{fig:functionalTopologyCentralisedApproach}).
%
%\begin{figure}[t]
%	\centering
%	\includegraphics[scale=0.35]{pics/centralisedApproach}
%	\caption{Functional topology that represents the centralised approach to assign frequencies}
%	\label{fig:functionalTopologyCentralisedApproach}
%\end{figure}

%% file: sections/complexity_model_section_nicked.tex
% !TeX spellcheck = en_GB
The traditional reductionist approach attempts to explain an entire system in terms of its individual components. In contrast, complex systems analysis is based on the relations between system parts which result in a greater outcome than would be expected from a simple sum of the outcomes of the individual parts. Therefore, complex systems move the focus from the node to the network.

As we represent network functions with topologies, the complex systems approach allows us to analyse the relations between functional parts. Such analysis captures the joint effort of functionally interconnected system parts and provides a measurement of the deviation of the complex behaviour compared to a linear (non-complex) system. In order to capture this deviation, we analyse the joint effort of system sub-parts which are represented with subgraphs of the functional topology. Different subgraphs with the same size capture the variety of organisational structures in the topology, whereas different subgraph sizes allow us to understand the gain from increasing number of nodes and interactions. 

In order to capture the non-linear joint effort of functional parts we analyse the interactions of nodes involved in the decision making process on different scales. The scale size $r$ determines the maximum number of allowed hops between two nodes in order to reach each other. In modelling a network function, our focus from the complexity point of view is the degree of interconnectivity which is represented by the reachability of functional parts. $R$ is the maximum scale size of the system, which is defined as the longest shortest path in the entire topology (i.e. the graph diameter of the functional topology). If $r = R$ the reachability of every node in the topology is equal to 1.

%\begin{table}
%		\caption{}
%  \label{equation_notions}
%
%  \vskip 1mm
%  \begin{tabular}{c|p{\dimexpr 0.6\linewidth-2\tabcolsep}}
%    \textbf{Symbol} & \textbf{Meaning}    \\ \cline{1-2}
%    \centering{$n$} & node
%    \\ \cline{1-2}
%    \centering{$N$} & total number of nodes in the functional topology
%    \\ \cline{1-2}
%    \centering{$j$} & subgraph size - number of nodes in the subgraph
%    \\ \cline{1-2}
%    \centering{$r$} & scale size
%    \\ \cline{1-2}
%    \centering{$R$} & maximum scale size, which is defined as the
%    longest shortest path in the whole functional topology
%    \\ \cline{1-2}
%    \centering{$i_r^n$} & number of nodes that can reach node $n$ for
%    a given subgraph
%    \\ \cline{1-2}
%    \centering{$x_n$} & Bernoulli random variable. $x_n = 1$ indicates
%    that an interaction in the course of function operation involves
%    node $n$, whereas $x_n=0$ indicates that the interaction does not
%    involve node $n$, for a given scale
%    \\ \cline{1-2}
%    \centering{$p_r(x_n=1), p_r(x_n=0)$} & probabilities that any
%    given interaction in the course of function operation involves or
%    does not involve node $n$ for a given scale r
%    \\ \cline{1-2}
%    \centering{$H_r(x_n)$} & entropy of node $n$ which indicates the
%    uncertainty of involvement of node $n$ in the operation of a
%    network function for a given scale size $r$
%    \\ \cline{1-2}
%    \centering{$\Lambda_k^j$} & $k^{th}$ subgraph with $j$ nodes 
%  \end{tabular}
%\end{table}

\begin{table}
	\caption{The notation used in the equations. }
	\label{equation_notions}
	
	\vskip 1mm
	\begin{tabular}{cp{\dimexpr 0.6\linewidth-2\tabcolsep}}
		\toprule
		\textbf{Symbol} & \textbf{Meaning}    		\\[0.9ex]  \midrule
		\centering{$n$} & node
		\\[0.9ex] 
		\centering{$N$} & total number of nodes in the functional topology
		\\[0.9ex] 
		\centering{$j$} & subgraph size - number of nodes in the subgraph
		\\[0.9ex] 
		\centering{$r$} & scale size
		\\[0.9ex] 
		\centering{$R$} & maximum scale size, which is defined as the
		longest shortest path in the whole functional topology
		\\[0.9ex] 
		\centering{$\Lambda_k^j$} & \textcolor{lastUpdate}{one of the $k$ subgraphs with $j$ nodes that is induced from the functional topology graph}
		\\[0.9ex]
		\centering{$i_r^n$} & number of nodes that can reach node $n$ for
		a given subgraph
		\\[0.9ex] 
		\centering{$x_n$} & Bernoulli random variable. $x_n = 1$ indicates
		that an interaction in the course of function operation involves
		node $n$, whereas $x_n=0$ indicates that the interaction does not
		involve node $n$, for a given scale
		\\[0.9ex] 
		\centering{$X$} & since $x_n$ is a Bernoulli random variable $X = \{0,1\}$
		\\[0.9ex] 
		\centering{$p_r(x_n=1), p_r(x_n=0)$} & probabilities that any
		given interaction in the course of function operation involves or
		does not involve node $n$ for a given scale r
		\\[0.9ex] 
		\centering{$H_r(x_n)$} & entropy of node $n$ which indicates the
		uncertainty of involvement of node $n$ in the operation of a
		network function for a given scale size $r$
		\\[0.9ex] 
		\bottomrule
	\end{tabular}
\end{table}

In order to capture the relationships that underpin the operation of a given network function, we consider the interactions enabled by the structure of the functional topology on various scales of operation. Specifically, we consider that a node interacts if another node can reach this node on a particular scale. As such, we employ the Bernoulli random variable $x_n$ to describe the potential for a node to interact on a particular scale, under the assumption that interactions are uniformly initiated by all nodes within the topology. Therefore, the probability distribution of $x_n$ is determined by the reachability of node $n$ for a given scale size $r$. Equation (\ref{eq:probabilityOfInteraction}) provides the relative reachability of node $n$, where $i_r^n$ is the number of nodes that can reach node $n$ for a given scale $r$ and $j$ represents the number of nodes for the given subgraph. Table \ref{equation_notions} summarises the notation used in our equations.

\begin{eqnarray}\label{eq:probabilityOfInteraction}
	p_r(x_n=1) = \dfrac{i_r^n}{j}
\end{eqnarray}

%\begin{equation}\label{eq:probabilityOfNotParticipation}
%	p_r(x_n=0) = \dfrac{(j-i_r^n)}{j}
%\end{equation}

\textcolor{lastUpdate}{Shannon entropy for each node of the given subgraph with size $j$, and scale size $r$ is calculated with equation (\ref{eq:entropy}).
\begin{eqnarray}\label{eq:entropy}
H_r(x_n) = \displaystyle\sum_{x_n \in X} p_r(x_n) \cdot \log_{2}\dfrac{1}{p_r(x_n)}
\end{eqnarray}}

%TODO what about values between 0 and 1 for the probabilities
%Considering that Shannon entropy is the average amount of information for a given node $n$ and 
\textcolor{lastUpdate}{Since the probabilities in equation (\ref{eq:probabilityOfInteraction}) and (\ref{eq:entropy}) indicate the relative reachability of a node, entropy in our functional model represents the uncertainty of interaction of node $n$ during the operation of a network function for a given scale size $r$.} The uncertainty of interaction of a node depends on the role of this node in the graph that represents the function (e.g. hub, stub, disconnected node). \textcolor{lastUpdate}{A node with zero entropy functionally represents a hub or a disconnected node.} A hub in the functional topology is a node connected to all other nodes, which means that the reachability of this node is $p_r(x_n=1) = 1$. Conversely, a disconnected node has the non-reachability $p_r(x_n=0) = 1$. \textcolor{lastUpdate}{A hub is a functionality or informational source that interacts with each subtask in the functional topology.} In contrast, a disconnected node in the functional framework represents a functionality not related to the modelled function, which means that this node does not provide information of interest for any part of our model. 
%As we analyse only connected graphs, $H_r(x_n) = 0$ means that node $n$ is a hub which can be reached from any other node for a given $r$.
%If $r = R$ the probabilities are $p_R(x_n=1) = 1$ and $p_R(x_n=0) = 0$ for every node $n$, which results in $H_R(x_n) = 0$ $\forall n$. 

The total amount of information of the $k^{th}$ subgraph with $j$ nodes for scale $r$ is calculated with equation (\ref{eq:totalAmountOfInformationSubgraph}). $\Lambda_k^j$ is one of the $k$ subgraphs with $j$ nodes. The total amount of information represents the total uncertainty which is related to the actual roles of nodes that appear within a subgraph and different subgraph patterns.\\

\begin{eqnarray}\label{eq:totalAmountOfInformationSubgraph}
I_r(\Lambda_k^j) = \displaystyle\sum_{n \in \Lambda_k^j} H_r(x_n)
\end{eqnarray}

The average amount of information for a given subgraph size $j$ is calculated with equation (\ref{eq:averageAmountOfInformationSubgraph}). $\beta_j$ in the equation represents the number of connected subgraphs with size $j$. 

\begin{eqnarray}\label{eq:averageAmountOfInformationSubgraph}
\langle I_r(\Lambda^j) \rangle = \dfrac{1}{\beta_j}\displaystyle\sum_{k =1}^{\beta_j} I_r(\Lambda_k^j)
\end{eqnarray}

Functional complexity is calculated with equation (\ref{eq:complexity}). 

\begin{eqnarray}\label{eq:complexity}
C_F = \dfrac{1}{R-1}\displaystyle\sum_{r =1}^{R-1}\sum_{j = 1+r}^{N} | \langle I_r(\Lambda^j) \rangle - \dfrac{r+1-j}{r+1-N}  I_r(\Lambda^N)|
\end{eqnarray}

Note that $\Lambda^N$ is the set of all nodes in the functional topology. For $j < 1 + r$ the amount of information is equal to zero, because every subgraph with size $j < 1 + r$ for the scale size $r$ represents a full mesh topology in terms of reachability, therefore $H_r(x_n) = 0$ and $I_r(\Lambda_k^j) = 0$. 

In order to determine the complexity of an implementation of a network function we have to compare this particular implementation to a non-complex model of itself. A non-complex model assumes that every functional part always contributes the same amount of information. \textcolor{lastUpdate}{Describing the functional parts provides full information about such a network function because the function is the sum of functionalities of its parts. Conversely, a complex implementation implies that the network function relies on communication between its parts which results in a higher utility outcome, i.e. greater than the sum of its parts. Complexity captures the variety of roles that each node has and the variety of structural patterns in the functional topology.}

The variety of roles emphasises the importance of interconnectivity of functional parts. Since nodes represent functional parts, different roles refer to different interactions that underpin the operation of a network function. If one node is highly connected to a group of nodes, this means that it influences the entire group. At the same time the group has a high influence on the operation of this node. In our frequency allocation example, the disconnected topology shows that every node is independent, which results in one simple role (disconnected node) for every subgraph. Describing the relationship between a single element and the rest of the functional topology provides all necessary information needed to understand the relationships of the entire topology. Therefore this structure is not considered as complex. Conversely, the self-organising algorithm provides different roles for nodes. This results in a complex structure, where the outcome depends on the communication between the structural parts, which means that the communication between system entities becomes more important to understand than the entities themselves.

%% file: sections/analysisNew_section.tex
% !TeX spellcheck = en_GB
\textcolor{lastUpdate}{In this section we apply our complexity model to several functional topologies to examine the main properties of the functional complexity metric.} In addition to the two implementations of the frequency allocation function, we also investigate common graph structures (e.g. ring, star, bus, full mesh) that describe complex and non-complex functional relations. In the course of this analysis, we present the relationship between the complexity of a function and graph theory metrics that describe the functional topology representation of the network function.

%A complex system, and hence a complex function, is determined by many parameters. Graph theory traditionally provides information about overall structural patterns. Complex systems analysis provides additional information about the relationships between system parts. We emphasize the importance of recognising several graph theory metrics that lead to certain values of complexity. These metrics are: 
A complex system, and hence a complex function, is determined by many parameters. We compare our metric to the following graph theoretic notions to highlight the utility of our methodology:
\begin{itemize}
	\item number of nodes,
	\item average path length,
	\item clustering coefficient.
\end{itemize}

\begin{figure}
	\includegraphics[scale=0.45]{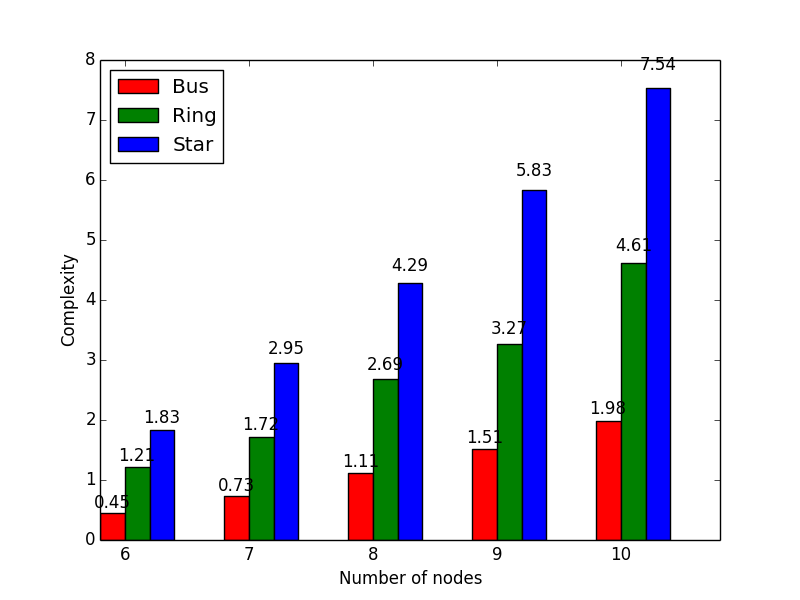}\vskip3mm
	\caption{\textcolor{activecolor}{Complexity of a bus, ring and star topology.}}
	\label{complexity_bus_ring_satr}
\end{figure}

Intuitively, if the number of nodes increases the system becomes more complex. The average path length indicates the distance between elements, which implicitly refers to the relationship strengths in the network. The clustering coefficient provides information about the overall strengths of functional dependencies between nodes grouped around a central entity. All these metrics provide useful information about certain organisational characteristics of the system. \textcolor{lastUpdate}{We start the analysis with common graph structures like a bus, ring, star and full mesh topology to investigate the impact of different graph structures on the amount of complexity.} Figure \ref{complexity_bus_ring_satr} shows complexity values of theses topologies in the range of six to ten nodes. Additionally, we present the correlation between the complexity metric and different graph theory metrics. 
%In contrast, complexity provides a holistic organisation degree analysis related to the structure of a network functions. 

%\begin{table}
%			\caption{Complexity of a bus, ring, star and full mesh topology}
%			\label{complexity_bus_ring_satr}
% 	\begin{tabular}{l|l|l|l|l|ll}
% 		& \multicolumn{5}{c}{Complexity}             \\ \cline{2-6}
% 		Topology & 6 nodes & 7 nodes & 8 nodes & 9 nodes & 10 nodes & \\ \cline{1-6}
% 		Bus           &   0.45         & 0.73 &  1.11 & 1.51  &  1.98 & \\ \cline{1-6}
% 		Ring            &   1.21         & 1.72 & 2.69 & 3.27  &  4.61 & \\ \cline{1-6}
% 		Star          &   1.83         &  2.95 &  4.29 &  5.83 & 7.54  & \\ \cline{1-6}
% 		Full mesh          &   0.00         &  0.00 & 0.00  & 0.00   & 0.00    &
% 	\end{tabular}
%\end{table}
%TODO caption change add more information
%TODO change the lable in the text Table -> Figure
\subsection{Bus, ring, star and full mesh topologies}

A full mesh topology results in zero complexity for any number of nodes. Therefore, even though a full mesh topology provides a densely connected structure, the overall relationships between functional parts represented by a full mesh topology are non-complex. The structures of any subgraph in a full mesh topology are also full mesh connectivity patterns. This means that a full mesh topology represents a function in which every functional part interacts with all other entities in order to make a decision. The functional structures and roles of nodes in a graph that represents such functions are always the same (each individual node is a hub). This means that in order to describe the system (the function), it is enough to describe an individual element and such functions are considered as non-complex.

\textcolor{lastUpdate}{All the nodes in a bus topology are fairly spread out (i.e. the average path length between nodes is high).} In other words, the bus topology represents a connectivity pattern with weak overall connections among parts of the topology. The bus topology does not provide a great variety of organisational structures, which results in low complexity. According to Figure \ref{complexity_bus_ring_satr}, functional complexity of the bus topology increases with the increase in the number of nodes. 
%Increasing number of nodes for a bus topology causes an increase in the average path length. The increase in the average path length weakens the relationship strengths of the topology and results in a decentralised structure. 

Figure \ref{complexity_bus_ring_satr} shows that the functional complexity of a ring topology is higher than the complexity of a bus topology for the same number of nodes. The ring topology conforms to the same trend in which complexity increases with the number of nodes. \textcolor{lastUpdate}{The ring topology compared to the bus is less spread out (i.e. the average path length is shorter) and therefore provides tighter connections between the nodes.} As the nodes are closer to each other, due to the additional path, this functional topology represents tighter functional relationships compared to the bus topology. Tighter functional relationships result in higher complexity.
%An obvious fact is that the average path length of the ring topology is shorter compared to the bus topology. This suggests tighter functional relationships because the nodes are closer to each other. Tighter functional relationships result in higher complexity of the ring compared to the bus topology. 

\textcolor{lastUpdate}{The star topology is a functional relationship in which one node interacts with all other nodes in the decision making process. It also conforms to the same trend in which complexity increases with the number of nodes in the topology. The star structure has tight connections between its parts. This means that the star topology depicts stronger functional relationships compared to the bus and ring topology. Therefore, the star topology has the highest complexity value compared to the other structures presented in Figure \ref{complexity_bus_ring_satr}.}
%The star structure has a very short average path length which indicates tighter functional relationships. 

%The analysis of different simple graph structures (bus, ring, star and full mesh) gives a general overview of the meaning of different graph theory metrics in terms of complexity. Considering the highly centralised functional relationship between entities represented with a star topology, we believe that this type of relationship contributes to the increase in complexity. Based on the analysis of different sizes of ring and bus topologies we believe that the increase in the average path length also affects the increasing complexity values.  
The analysis of simple graph structures (bus, ring, star and full mesh) gives a general overview of the impact of different graph organisations on the amount of complexity. The functional complexity metric captures the diversity of structures and roles of nodes, which provides a different perspective to analyse the complex relationships between system entities. According to Figure \ref{complexity_bus_ring_satr}, the star topology is the most complex structure compared to the bus, ring and full mesh topologies. \textcolor{activecolor}{Considering the sparse connectivity and short average path length between entities represented with a star topology, we believe that this type of relationship contributes to the increase in complexity.} Based on the analysis of different sizes of ring and bus topologies, we believe that spreading (distributing) the functionality all over the topology also affects the increasing complexity values. We expect that a combination of these two types of relationships results in highly complex organisational structures. Such a structure consists of local centres (star subgraphs/clusters) that are highly interconnected. As functional complexity captures the variety of structural patterns, we expect that complexity is high if the number of local centres is high. 

\subsection{Comparison with graph theory metrics}

\begin{figure}
	\includegraphics[scale=0.45]{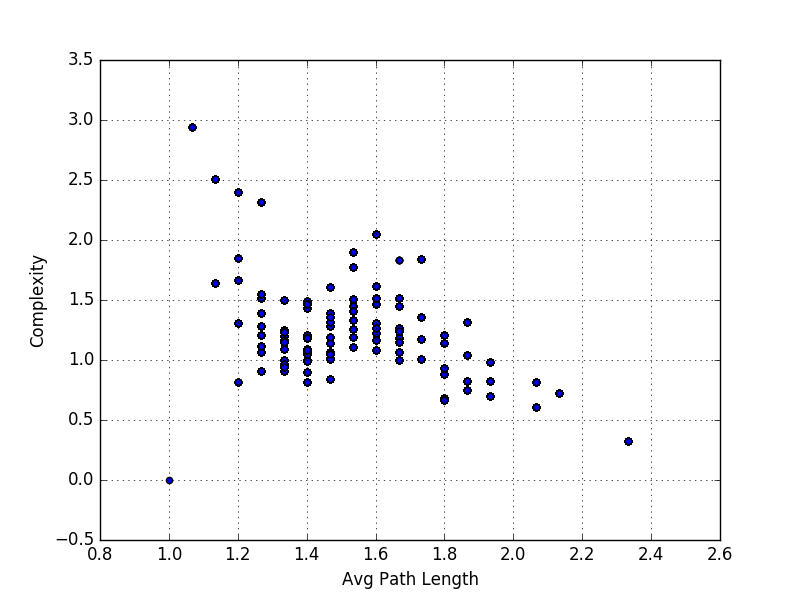}\vskip3mm
	\caption{\textcolor{activecolor}{The relationship between the average path length and functional complexity for all possible distributions of links between six nodes. The Pearson product-moment correlation coefficient for these two variables (average path length and functional complexity) is -0.43.}}
	\label{avg_path_len_complx}
\end{figure}

To describe the relationship between the average path length and the functional complexity of an implementation of a network function we analyse all possible graph organisations (distributions of links) between six nodes - see Figure \ref{avg_path_len_complx}. \textcolor{activecolor}{The Pearson product-moment correlation coefficient for these two metrics is -0.43 (see Table \ref{correlation_table})}. Notice in Figure \ref{avg_path_len_complx} that the highest functional complexity for six nodes is 2.9. Compared to the bus, ring and star topology the highest functional topology is at least 1.5 times higher. An important fact is that for the same average path length we have a range of different complexity values. Therefore, despite the high correlation between the average path length and complexity, we can't predict the complexity of a structure based on the average path length. Additionally, the maximum value of complexity occurs for an extremely low average path length, which is still higher than one. Figure \ref{avg_path_len_complx} also shows that the envelope of the complexity values is a non-monotonic function. That is, there exist complexity values for longer average path lengths that are larger than all complexity values for some shorter average path lengths. Notice also that complexity is equal to zero for the average path length equal to one. This value of the average path length represents a full mesh topology.  

\begin{figure}
 	\includegraphics[scale=0.45]{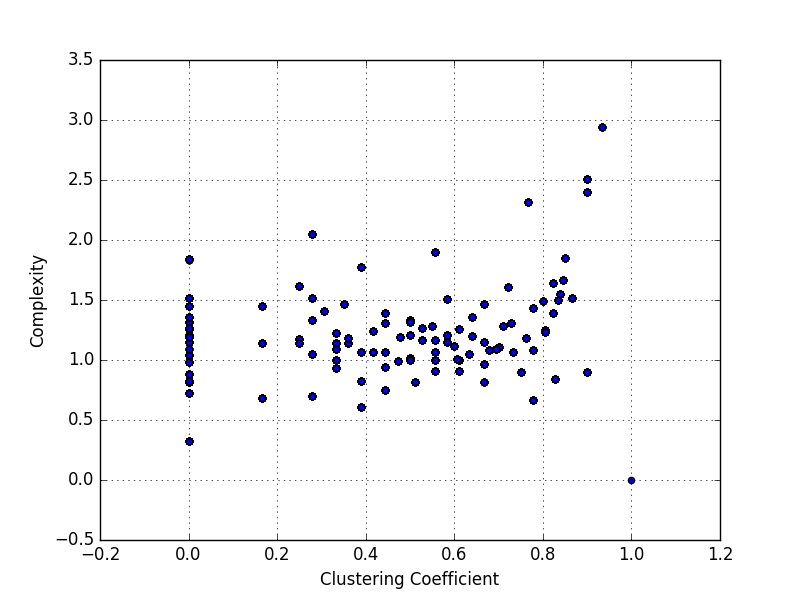}\vskip3mm
 	\caption{\textcolor{activecolor}{The relationship between the clustering coefficient and the functional complexity for all possible distributions of links between six nodes. The Pearson product-moment correlation coefficient for these two variables (clustering coefficient and functional complexity) is 0.15.}}
 	\label{clustering_complx}
 \end{figure}

As shown in Table \ref{correlation_table}, the correlation between the clustering coefficient and complexity is low (0.15) for six nodes. Figure \ref{clustering_complx} depicts the relationship between the average clustering coefficient and complexity for six nodes. The average clustering coefficient is a graph theory metric that measures the average number of neighbouring nodes that are neighbours to each other \cite{Newman2010}. Therefore, the average clustering coefficient quantifies the strength of functional connections between nodes grouped around local centres. Again, the envelope of the complexity values in this relationship is also non-monotonic. Notice in Figure \ref{clustering_complx} that complexity has its maximum for an average clustering coefficient close to one. This is expected because this value of the clustering coefficient represents a network with tight functional connections between nodes grouped around local centres. Notice also that complexity is zero for the average clustering coefficient equal to one. According to the definition of the clustering coefficient, the value one indicates that all neighbours of every node in the functional topology are neighbours to each other, which represents a full mesh topology. 

\begin{figure}
  	\includegraphics[scale=0.45]{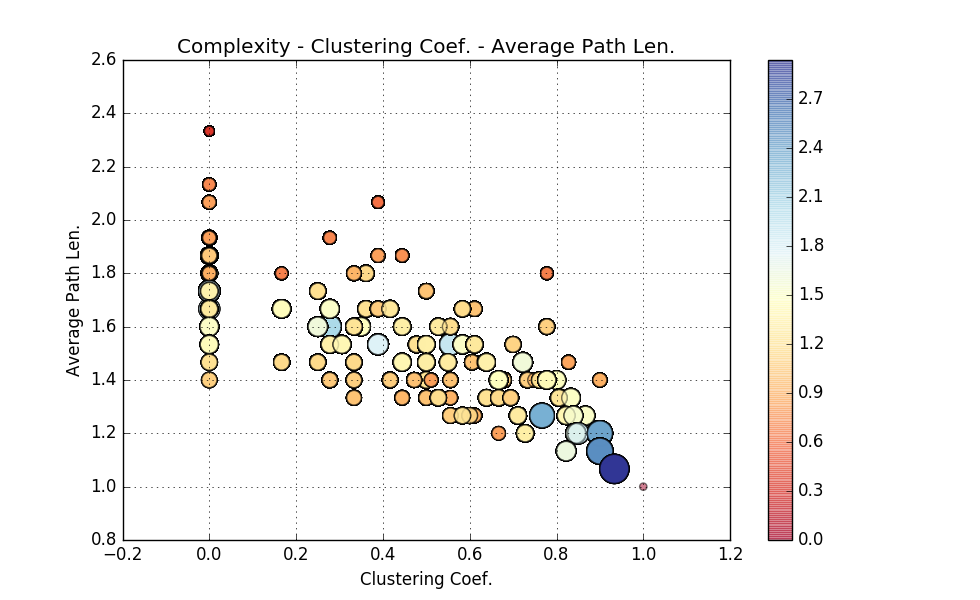}\vskip3mm
  	\caption{\textcolor{activecolor}{The relationship between the clustering coefficient, average path length and the functional complexity for all possible distributions of links between six nodes. The complexity value is represented by the colour and diameter of circles.}}
  	\label{clustering_avg_path_len_complx}
\end{figure}

In order to analyse the joint relationship between complexity and both of the graph theory metrics presented above (the average path length and the average clustering coefficient) we plot all these metrics together on one graph (see Figure \ref{clustering_avg_path_len_complx}). The x and y axis represent the average clustering coefficient and the average path length respectively and the complexity is represented by the radius of the circles. Additionally, the circles are coloured differently to emphasize the different complexity values. The multivariate correlation between these three variables is 0.47 (see Table \ref{correlation_table}). Notice that we have overlapping circles with different radii on the graph which means that for the same average path length and clustering coefficient we observe functional topologies with different complexities. As shown in the analysis above, graph theory metrics allow us to analyse individual aspects of the functional topology, whereas complexity adds additional information about the diversity of relationships between functional parts and provides a different approach to analyse complex network functions. 
 
%\begin{table}
%	\centering
%	\caption{Correlation between graph theory metrics and the functional complexity metric for all possible distributions of links between six nodes.}
%	\label{correlation_table}
%	\begin{tabular}{l|llll}
%		Correlation variables & Correlation &  &  &  \\ \cline{1-2}
%		Average Path length - Complexity            &    -0.43        &  &  &  \\ \cline{1-2}
%		Average Degree Distribution - Complexity  &    0.29       &  &  &  \\ \cline{1-2}
%		Clustering coefficient - Complexity            &    0.15        &  &  &  \\ \cline{1-2}
%		Avg. Path length - Clustering coef. - Complexity          &    0.47        &  &  &  \\ \cline{1-2}
%		Avg. Path length - Average Degree Dist. - Complexity          &    0.49        &  &  &  \\ \cline{1-2}
%		Clustering coef. - Average Degree Dist. - Complexity          &    0.30       &  &  &  
%	\end{tabular}
%\end{table}

\begin{table}
	\centering
	\caption{Correlation between graph theory metrics and the functional complexity metric for all possible distributions of links between six nodes.}
	\label{correlation_table}
	\begin{tabular}{lp{\dimexpr 0.3\linewidth-10\tabcolsep}}
		\toprule
		Correlation variables & Correlation \\[0.9ex]
		\midrule
		Average Path length - Complexity            &    -0.43       \\[0.9ex] 
%		Average Degree Distribution - Complexity  &    0.29    \\[0.9ex]
		Clustering coefficient - Complexity            &    0.15        \\[0.9ex]
		Avg. Path length - Clustering coef. - Complexity          &    0.47     \\[0.9ex]
%		Avg. Path length - Average Degree Dist. - Complexity          &    0.49        \\[0.9ex]
%		Clustering coef. - Average Degree Dist. - Complexity          &    0.30  \\[0.9ex]
		\bottomrule
	\end{tabular}
\end{table}

\begin{figure}
	\includegraphics[scale=0.45]{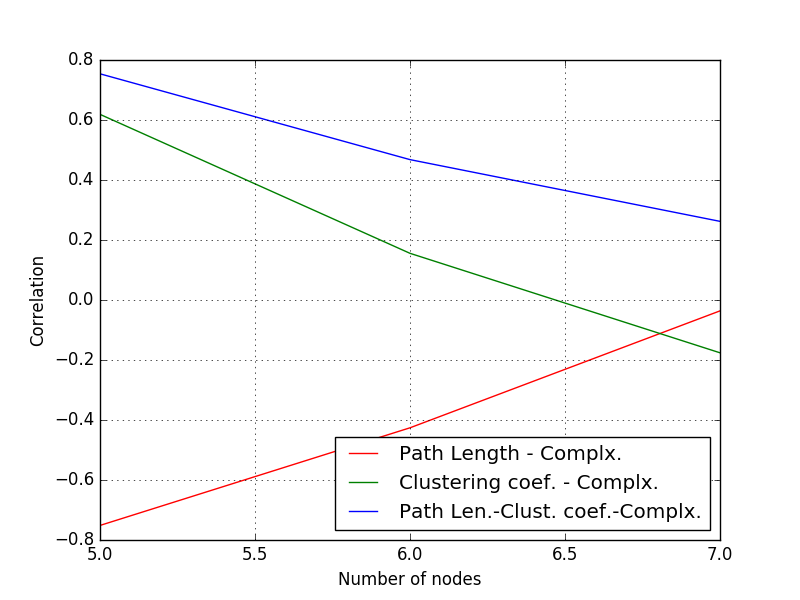}\vskip3mm
	\caption{\textcolor{activecolor}{The relationship between the correlation coefficients of the graph theory metrics (average path length, clustering coefficient) and the complexity metric for different sizes of functional topologies.}}
	\label{three_lines_graph}
\end{figure}

Figure \ref{three_lines_graph} shows that  the correlation between complexity and different graph theory metrics decreases as the number of nodes in the graph increases. This is expected, because the number of graph combinations increases exponentially with the number of nodes. The rapid increase of graph combinations allows us to create a variety of complex structures for the same value of certain graph theory metrics. Hence, the complexity range increases for an individual value of a graph theory metric. The decrease of correlation between complexity and different graph theory metrics suggests that graph theory metrics do not target the same properties of the system. This means that the complexity metric adds new information, which leads to better understanding of the relationships between system entities. 

%TODO CHANGE CAPTION check text in paragraphs

\subsection{Complexity of the frequency allocation algorithms}

In order to analyse the functional complexity of the two frequency allocation algorithms presented in the framework section we apply our complexity model to the functional topology representation of these network functions. Our complexity metric, when applied over the functional topology, gives insight into the functional relationships between system entities, which allows us to focus on the joint effect of functional parts in order to execute the modelled network function.

The random frequency allocation algorithm is represented by a disconnected graph. Each element represents a closed system which results in the absence of connections between them. From the system perspective the functional parts of this implementation represent a set of independent elements. As the elements do not interact with each other, in order to describe the set we can simply describe its elements. More precisely, to describe the set we have to describe only one element, because each element of the set is the same. Each element of the set simply assigns one of the available frequencies. Such a set that is represented with independent elements has zero functional complexity.

\begin{figure}
	\centering \subfigure[\textcolor{activecolor}{The relationship between the uncertainty of interactions for all subset sizes and the uncertainty which is
	expected from the calculation performed on the whole system.}]{
		\includegraphics[scale=0.32]{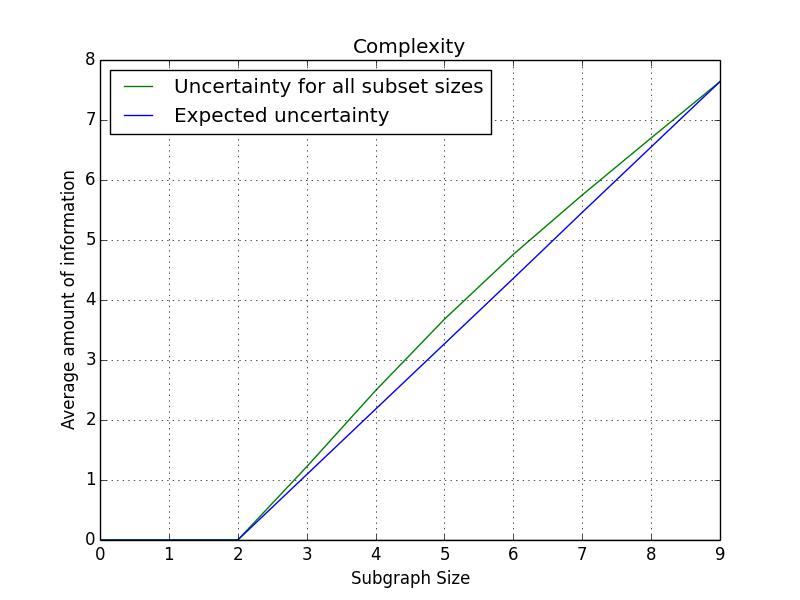}
		\label{complexity_self_organized_alg} } 
	\hskip9mm \subfigure[\textcolor{activecolor}{The distance between the uncertainty of interactions for all subset sizes and the uncertainty which is expected from the calculation performed on the whole system, i.e. the functional complexity of the self-organising frequency assignment approach.}]{
		\includegraphics[scale=0.32]{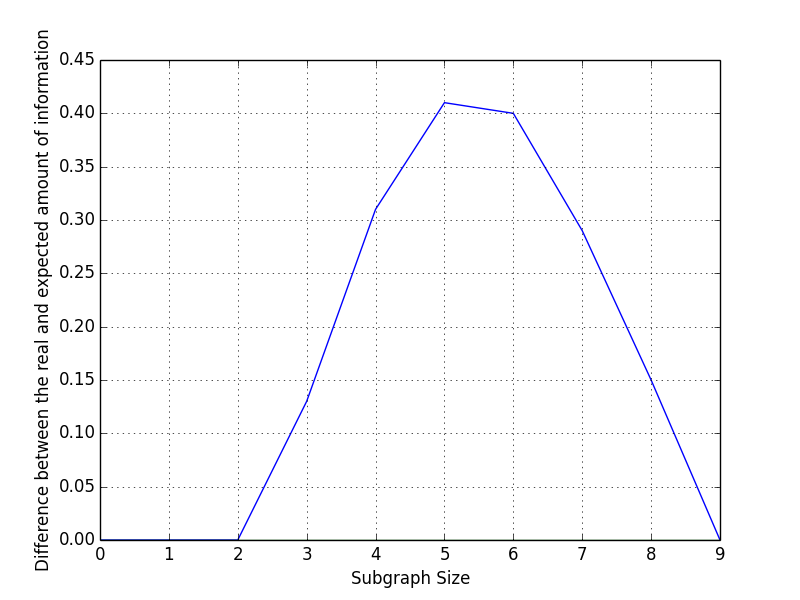}
		\label{real_vs_expected_inf} }
	
	\caption{\textcolor{activecolor}{The functional topology has nine nodes; the maximum scale size $R$ is 2; the functional complexity is represented by the area between the linear and non-linear function.}}
	\label{complexities}
\end{figure}

Finally, the self-organising distributed frequency allocation algorithm is represented by a functional topology shown in Figure \ref{fig:funcTopologySelfOrganising}. In order to analyse the complexity of this implementation of the frequency allocation function we apply our complexity model to the functional topology representation. Figure \ref{complexity_self_organized_alg} depicts the deviation in the amount of information with increasing subgraph sizes from a linear increase. The linear increase represents an equivalent simple system implementation, meaning that all the subgraphs of a specific size feature the same uncertainty of interaction. Every functional part contributes with the same amount of information, which results in a linear increase over increasing subgraph sizes. The non-linear increase represents the heterogeneity of structures within the functional representation of a network function, which indicates that the relationship between functional parts is complex. Considering that for the functional topology of the self-organising algorithm $R=2$, the multi-scale approach applied on this functional topology implies only a single scale analysis, that is $r=1$. The subgraph sizes of interest are in the range of two to nine. Considering that $r=1$, subgraphs with the size two or smaller represent full mesh topologies. The maximum value of the subgraph size is nine because the functional topology consists of nine nodes. The complexity of the self-organising frequency allocation algorithm is 1.69 and it is represented by the area between the linear and non-linear functions in Figure \ref{complexity_self_organized_alg}. \textcolor{activecolor}{Figure \ref{real_vs_expected_inf} shows the distance between the uncertainty of interactions for all subset sizes and the uncertainty which is expected from the calculation performed on the whole system, i.e. the functional complexity of the self-organising frequency allocation approach. More precisely, Figure \ref{real_vs_expected_inf} depicts the single scale complexity function which is represented by the inner sum of equation (\ref{eq:complexity}). The functional complexity is the sum of the function depicted in Figure \ref{real_vs_expected_inf}.}

%\begin{figure}
%	\includegraphics[scale=0.45]{pics/ComplexityGraph}\vskip3mm
%	\caption{Functional complexity of the self-organising frequency assignment approach; the functional topology has nine nodes; the maximum scale size $R$ is 2; the functional complexity is represented by the area between the linear and non-linear function.}  
%	\label{complexity_self_organized_alg}
%%	\vspace{-5mm}
%\end{figure}

%TODO comment on graph b
Similarly, the authors of \cite{Macaluso2016} analyse the two frequency allocation algorithms. In contrast to our work, they analyse the complexity of the outcome (i.e. the frequency allocation) of these implementations. The results that are presented in their work show that the random approach produces an outcome with zero complexity\footnote{The complexity metric used in \cite{Macaluso2016} is excess entropy.}, whereas the self-organising distributed frequency allocation algorithm produces a complex outcome. Here, our functional complexity metric analyses the complexity of the implementation itself. As discussed above, the functional complexity of the random frequency allocation algorithm is zero, and the functional complexity of the self-organising distributed frequency allocation algorithm is 1.69. This suggests that a complex implementation of a function results in a complex outcome. In \cite{Macaluso2016} the authors show that the response to a change of a system with highly complex output happens in a more manageable fashion causing less disruption. The authors of \cite{Macaluso2016} also emphasize the higher robustness of a complex outcome compared to a non-complex outcome. Making a connection between the implementation and the outcome of a function correlates the characteristics of the outcome with the complex relationships that underpin the functional structure.

%% file: sections/conclusion_section.tex
% !TeX spellcheck = en_GB
The growing size and heterogeneity of telecommunication networks leads to a need to change the way we analyse and model them. Also the rapid evolution of network services demands a new approach to analyse them. The aim of this paper was to contribute to this new way of studying networks. We focus on network functions as building blocks of services. We consider network functions as complex systems. In order to provide a new approach to analyse and understand the impact of complex functional relationships between system entities, we developed a new framework. The framework allows us to visualise an implementation of a network function with graphs called functional topologies. We provide several examples that show how to map a network function into a functional topology. The graph visualisation of the implementation allows us to focus on the relationships between entities rather than the entities themselves. 

The next step after mapping an implementation of a network function into a functional topology was to provide a metric that quantifies the organisational structure of the topology. Our functional complexity ($C_F$) captures the variety of roles that each node in the topology has and the variety of structural patterns present in the topology. $C_F$ quantifies the deviation of the implementation of a network function from the non-complex model of itself. The quantification of this deviation as presented in Section \ref{sec:complexityModel} provides a new approach to understand increasingly complex telecommunication networks. 

In order to study the impact of different structural patterns on the functional complexity in Section \ref{sec:Analysis}, we start by analysing the complexity of simple graph structures (bus, ring, star and mesh). Additionally, we provide a detailed study that investigates the impact of several graph theory metrics on the functional complexity. \textcolor{lastUpdate}{We investigated the correlation between the combination of the graph theory metrics and complexity, due to the absence of high correlation between any graph theory metrics and the functional complexity.} This analysis allowed us to make conclusions about the organisational structures that is needed in order to achieve high complexity.

In this paper we also analysed the functional complexity of two different implementations of the frequency allocation function (random and self-organising). First, in Section \ref{sec:framework} we explain how to map these implementations into functional topologies, and in Section \ref{sec:Analysis}, we apply our complexity metric to quantify the functional complexity of these implementations. We showed that the random frequency allocation has zero complexity (and represents therefore a non-complex implementation), whereas the self-organising implementation shows higher complexity. This allowed us to make a connection between our results and the results from \cite{Macaluso2016}, where the authors calculated the complexity of the outcome of these implementations and got the same kind of results. This implies that a complex implementation results in a complex outcome of the function and allows us to make assumptions about the robustness and response to change similar to \cite{Macaluso2016}.

Overall, our complexity metric quantifies telecommunication networks in terms of their functional relationships and provides a wholly new approach to understanding the operation of networks. More precisely, the complexity metric quantifies the relationships employed during operation of network functions. This provides a new approach to study networks which is especially relevant for next generation networks.
%Complexity allows us to model the next generation networks in this manner. 

%The link between complexity, robustness and response to change suggests that complex system analysis applied on telecommunication networks can provide information about the underlying mechanisms that enable certain network characteristics. Therefore, c

%% file: sections/acronyms_section.tex
% !TeX spellcheck = en_GB
\begin{acronym}
	\acro{RNC}{Radio Network Controler}
	\acro{nodeB}{UMTS base transciever station}
	\acro{VLR}{Visitor Location Register} 
	\acro{HLR}{Home Location Register}
	\acro{SGSN}{Serving GPRS Support Node}
	\acro{MSC}{Mobile Switching Center}
	\acro{SMS}{Short Message Service}
	\acro{IoT}{Internet of Things}
	\acro{CPR}{Common-Pool Resource}
\end{acronym}

%% file: A_Functional_Complexity_Framework_for_the_Analysis_of_Telecommunication_Networks.bbl
\begin{thebibliography}{10}
\providecommand{\url}[1]{#1}
\csname url@samestyle\endcsname
\providecommand{\newblock}{\relax}
\providecommand{\bibinfo}[2]{#2}
\providecommand{\BIBentrySTDinterwordspacing}{\spaceskip=0pt\relax}
\providecommand{\BIBentryALTinterwordstretchfactor}{4}
\providecommand{\BIBentryALTinterwordspacing}{\spaceskip=\fontdimen2\font plus
\BIBentryALTinterwordstretchfactor\fontdimen3\font minus
  \fontdimen4\font\relax}
\providecommand{\BIBforeignlanguage}[2]{{%
\expandafter\ifx\csname l@#1\endcsname\relax
\typeout{** WARNING: IEEEtran.bst: No hyphenation pattern has been}%
\typeout{** loaded for the language `#1'. Using the pattern for}%
\typeout{** the default language instead.}%
\else
\language=\csname l@#1\endcsname
\fi
#2}}
\providecommand{\BIBdecl}{\relax}
\BIBdecl

\bibitem{Liu2013}
\BIBentryALTinterwordspacing
Y.-Y. Liu, J.-J. Slotine, and A.-L. Barab\'{a}si, ``{Observability of complex
  systems.}'' \emph{Proceedings of the National Academy of Sciences of the
  United States of America}, vol. 110, no.~7, pp. 2460--5, 2013. [Online].
  Available: \url{http://www.pnas.org/content/110/7/2460.full}
\BIBentrySTDinterwordspacing

\bibitem{Bar-Yam2004}
Y.~Bar-Yam, ``{Multiscale Complexity/Entropy},'' \emph{Advances in Complex
  Systems}, vol.~07, no.~1, pp. 47--63, 2004.

\bibitem{Yaeger2007}
\BIBentryALTinterwordspacing
L.~Yaeger, ``{Evolution Selects For and Against Complexity},'' no. November,
  2007. [Online]. Available:
  \url{http://vw.slis.indiana.edu/talks-fall07/Yeager.pdf}
\BIBentrySTDinterwordspacing

\bibitem{Tononi1994}
G.~Tononi, O.~Sporns, and G.~M. Edelman, ``{A measure for brain complexity:
  relating functional segregation and integration in the nervous system.}''
  \emph{Proceedings of the National Academy of Sciences of the United States of
  America}, vol.~91, no.~11, pp. 5033--5037, 1994.

\bibitem{Lopez-Ruiz1995}
R.~Lopez-Ruiz, H.~L. Mancini, and X.~Calbet, ``{A statistical measure of
  complexity},'' \emph{Physics Letters A}, vol. 209, no. 5-6, pp. 321--326,
  1995.

\bibitem{Balduzzi2008}
D.~Balduzzi and G.~Tononi, ``{Integrated information in discrete dynamical
  systems: Motivation and theoretical framework},'' \emph{PLoS Computational
  Biology}, vol.~4, no.~6, 2008.

\bibitem{Joshi2013}
\BIBentryALTinterwordspacing
N.~J. Joshi, G.~Tononi, and C.~Koch, ``{The Minimal Complexity of Adapting
  Agents Increases with Fitness},'' \emph{PLoS Comput Biol}, vol.~9, no.~7, p.
  e1003111, 2013. [Online]. Available:
  \url{http://dx.doi.org/10.1371/journal.pcbi.1003111}
\BIBentrySTDinterwordspacing

\bibitem{Evans2002}
T.~P. Evans, E.~Ostrom, and C.~Gibson, ``{Scaling Issues With Social Data in
  Integrated Assessment Modeling},'' \emph{Integrated Assessment}, vol.~3, pp.
  135--150, 2002.

\bibitem{Lloyd2001}
S.~Lloyd, ``{Measures of complexity: a nonexhaustive list},'' \emph{Control
  Systems, IEEE}, vol.~21, no.~4, pp. 7--8, 2001.

\bibitem{Wang2009}
P.~Wang, M.~C. Gonzalez, C.~A. Hidalgo, and A.-L. Barabasi, ``{Understanding
  the Spreading Patterns of Mobile Phone Viruses},'' \emph{Science}, vol. 324,
  no. May, pp. 1071--1076, 2009.

\bibitem{Onnela2007}
J.-P. Onnela, J.~Saram\"{a}ki, J.~Hyv\"{o}nen, G.~Szab\'{o}, D.~Lazer,
  K.~Kaski, J.~Kert\'{e}sz, and a~L~Barab\'{a}si, ``{Structure and tie
  strengths in mobile communication networks.}'' \emph{Proceedings of the
  National Academy of Sciences of the United States of America}, vol. 104,
  no.~18, pp. 7332--7336, 2007.

\bibitem{Macaluso2014}
I.~Macaluso, H.~Cornean, N.~Marchetti, and L.~Doyle, ``{Complex communication
  systems achieving interference-free frequency allocation},''
  \emph{Communications (ICC), 2014 IEEE International Conference on}, pp.
  1447--1452, 2014.

\bibitem{Macaluso2016}
I.~Macaluso, C.~Galiotto, N.~Marchetti, and L.~Doyle, ``{A Complex Systems
  Science Perspective on Wireless Networks},'' \emph{Journal of Systems Science
  and Complexity}, no.~10, pp. 1--27, 2016.

\bibitem{kaminski2014social}
N.~J. Kaminski, ``{Social Intelligence for Cognitive Radios},'' Ph.D.
  dissertation, Virginia Polytechnic Institute and State University,
  Blacksburg, Virginia, 2014.

\bibitem{Newman2010}
M.~Newman, \emph{Networks: An Introduction}.\hskip 1em plus 0.5em minus
  0.4em\relax New York, NY, USA: Oxford University Press, Inc., 2010.

\end{thebibliography}


\begin{thebibliography}{199}
\bibitem{Liu2013}
Y.-Y. Liu, J.-J. Slotine, and A.-L. Barab\'{a}si, ``{Observability of complex
	systems.}'' \emph{Proceedings of the National Academy of Sciences of the
	United States of America}, vol. 110, no.~7, pp. 2460--5, 2013. 

\bibitem{Bar-Yam2004}
Y.~Bar-Yam, ``{Multiscale Complexity/Entropy},'' \emph{Advances in Complex
	Systems}, vol.~07, no.~1, pp. 47--63, 2004.

\bibitem{Yaeger2007}
L.~Yaeger, ``{Evolution Selects For and Against Complexity},'' no. November,
2007. Available: http://vw.slis.indiana.edu/talks-fall07/Yeager.pdf

\bibitem{Tononi1994}
G.~Tononi, O.~Sporns, and G.~M. Edelman, ``{A measure for brain complexity:
	relating functional segregation and integration in the nervous system.}''
\emph{Proceedings of the National Academy of Sciences of the United States of
	America}, vol.~91, no.~11, pp. 5033--5037, 1994.

\bibitem{Lopez-Ruiz1995}
R.~Lopez-Ruiz, H.~L. Mancini, and X.~Calbet, ``{A statistical measure of
	complexity},'' \emph{Physics Letters A}, vol. 209, no. 5-6, pp. 321--326,
1995.

\bibitem{Balduzzi2008}
D.~Balduzzi and G.~Tononi, ``{Integrated information in discrete dynamical
	systems: Motivation and theoretical framework},'' \emph{PLoS Computational
	Biology}, vol.~4, no.~6, 2008.

\bibitem{Joshi2013}
N.~J. Joshi, G.~Tononi, and C.~Koch, ``{The Minimal Complexity of Adapting
	Agents Increases with Fitness},'' \emph{PLoS Comput Biol}, vol.~9, no.~7, p.
e1003111, 2013. 

\bibitem{Evans2002}
T.~P. Evans, E.~Ostrom, and C.~Gibson, ``{Scaling Issues With Social Data in
	Integrated Assessment Modeling},'' \emph{Integrated Assessment}, vol.~3, pp.
135--150, 2002.

\bibitem{Lloyd2001}
S.~Lloyd, ``{Measures of complexity: a nonexhaustive list},'' \emph{Control
	Systems, IEEE}, vol.~21, no.~4, pp. 7--8, 2001.

\bibitem{Wang2014}
Z.~Wang, D.~Zhang, X.~Zhou, D.~Yang, Z.~Yu, and Z.~Yu, ``{Discovering and
	profiling overlapping communities in location-based social networks},''
\emph{IEEE Transactions on Systems, Man, and Cybernetics: Systems}, vol.~44,
no.~4, pp. 499--509, 2014.

\bibitem{Ahn2010}
Y.-Y. Ahn, J.~P. Bagrow, and S.~Lehmann, ``{Link communities reveal multiscale
	complexity in networks.}'' \emph{Nature}, vol. 466, no. 7307, pp. 761--764,
2010.

\bibitem{Evans2009}
T.~S. Evans and R.~Lambiotte, ``{Line graphs, link partitions, and overlapping
	communities},'' \emph{Physical Review E - Statistical, Nonlinear, and Soft
	Matter Physics}, 2009.

\bibitem{Newman2003}
M.~E.~J. Newman and M.~Girvan, ``Finding and evaluating community structure in
networks,'' \emph{Phys. Rev. E}, vol.~69, no.~2, p. 026113, Feb. 2004.

\bibitem{Newman2006}
M.~E.~J. Newman, ``Modularity and community structure in networks,''
\emph{Proceedings of the National Academy of Sciences}, vol. 103, no.~23, pp.
8577--8582, 2006.

\bibitem{Lanham2013}
M.~J. Lanham, G.~P. Morgan, and K.~M. Carley, ``{Social Network Modeling and
	Agent-Based Simulation in Support of Crisis De-Escalation},'' \emph{IEEE
	Transactions on Systems, Man and Cybernetics Part C: Applications and
	Reviews}, vol.~44, no.~1, pp. 103--110, 2013.

\bibitem{Bristow2014}
M.~Bristow, L.~Fang, and K.~W. Hipel, ``{Agent-based modeling of competitive
	and cooperative behavior under conflict},'' \emph{IEEE Transactions on
	Systems, Man, and Cybernetics: Systems}, vol.~44, no.~7, pp. 834--850, 2014.

\bibitem{Gonzalez2008}
M.~Gonz\'{a}lez, C.~Hidalgo, and a.~L.~Barab\'{a}si, ``{Understanding
	individual human mobility patterns},'' \emph{Nature}, vol. 453, no. June, pp.
779--782, 2008.

\bibitem{Wang2009}
P.~Wang, M.~C. Gonzalez, C.~A. Hidalgo, and A.-L. Barabasi, ``{Understanding
	the Spreading Patterns of Mobile Phone Viruses},'' \emph{Science}, vol. 324,
no. May, pp. 1071--1076, 2009.

\bibitem{Onnela2007}
J.-P. Onnela, J.~Saram\"{a}ki, J.~Hyv\"{o}nen, G.~Szab\'{o}, D.~Lazer,
K.~Kaski, J.~Kert\'{e}sz, and a~L~Barab\'{a}si, ``{Structure and tie
	strengths in mobile communication networks.}'' \emph{Proceedings of the
	National Academy of Sciences of the United States of America}, vol. 104,
no.~18, pp. 7332--7336, 2007.

\bibitem{Beigy2010}
H.~Beigy and M.~R. Meybodi, ``{Cellular learning automata with multiple
	learning automata in each cell and its applications},'' \emph{IEEE
	Transactions on Systems, Man, and Cybernetics, Part B: Cybernetics}, vol.~40,
no.~1, pp. 54--65, 2010.

\bibitem{Macaluso2014}
I.~Macaluso, H.~Cornean, N.~Marchetti, and L.~Doyle, ``{Complex communication
	systems achieving interference-free frequency allocation},''
\emph{Communications (ICC), 2014 IEEE International Conference on}, pp.
1447--1452, 2014.

\bibitem{Macaluso2016}
I.~Macaluso, C.~Galiotto, N.~Marchetti, and L.~Doyle, ``{A Complex Systems
	Science Perspective on Wireless Networks},'' \emph{Journal of Systems Science
	and Complexity}, no.~10, pp. 1--27, 2016.

\bibitem{kaminski2014social}
N.~J. Kaminski, ``{Social Intelligence for Cognitive Radios},'' Ph.D.
dissertation, Virginia Polytechnic Institute and State University,
Blacksburg, Virginia, 2014.

\bibitem{Newman2010}
M.~Newman, \emph{Networks: An Introduction}.\hskip 1em plus 0.5em minus
0.4em\relax New York, NY, USA: Oxford University Press, Inc., 2010.
\end{thebibliography}
